\documentclass[prd,aps,showpacs,nofootinbib]{revtex4-1}%
\usepackage{amsmath}
\usepackage{amsfonts}
\usepackage{amssymb}
\usepackage{bm}
\usepackage{graphicx}
\usepackage{subfig}
\usepackage{color}
\usepackage{ulem}

\begin{document}
\title{Maxwell electrodynamics modified by \textit{CPT}-even and Lorentz-violating \\
dimension-6 higher-derivative terms}
\author{Rodolfo Casana$^{a}$}\email{rodolfo.casana@ufma.br}
\author{Manoel M. Ferreira,~Jr.$^{a}$}\email{manojr.ufma@gmail.com}
\author{Let\'{\i}cia Lisboa-Santos$^{a}$}\email{let\_lisboa@hotmail.com}
\author{Frederico E.P. dos Santos$^{b}$}\email{frederico.santos@ufma.br}
\author{Marco Schreck$^{a}$}\email{marco.schreck@ufma.br}
\affiliation{$^{a}$Departamento de F\'{\i}sica, Universidade Federal do Maranh\~{a}o
(UFMA), Campus Universit\'{a}rio do Bacanga, S\~{a}o Lu\'{\i}s -- MA,
65080-805, Brazil}
\affiliation{$^{b}$Coordena\c{c}\~{a}o do Curso Interdisciplinar em Ci\^{e}ncia e Tecnologia, Universidade Federal do Maranh\~{a}o (UFMA), Campus Universit\'{a}rio do Bacanga, S\~{a}o Lu\'{\i}s -- MA,
65080-805, Brazil}

\begin{abstract}
In this paper, we investigate an electrodynamics in which the physical modes are
coupled to a Lorentz-violating (LV) background by means of a higher-derivative term.
We analyze the modes associated with the dispersion relations (DRs) obtained from
the poles of the propagator. More specifically, we study Maxwell's electrodynamics
modified by a LV operator of mass dimension 6. The modification has the form
${D_{\beta\alpha}}\partial_{\sigma}F^{\sigma\beta}\partial_{\lambda}
F^{\lambda\alpha}$, i.e., it possesses two additional derivatives coupled to a
\textit{CPT}-even tensor $D_{\beta\alpha}$ that plays the role of the fixed background.
We first evaluate the propagator and obtain the dispersion relations
of the theory. By doing so, we analyze some configurations of the fixed background
and search for sectors where the energy is well-defined and causality is assured. A
brief analysis of unitarity is included for particular configurations.
Afterwards, we perform the same kind of analysis for a more general
dimension-6 model. We conclude that the modes of both Lagrange densities are possibly
plagued by physical problems, including causality and unitarity violation, and that signal
propagation may become physically meaningful only in the high-momentum regime.
\end{abstract}

\maketitle

\section{Introduction}

Physics beyond the Standard Model has been under extensive development in the latest
years, encompassing Lorentz-violating (LV) theories as one branch of investigation.
The minimal Standard-Model Extension (SME) \cite{Colladay,Samuel} is a general
gauge-invariant and power-counting renormalizable framework that incorporates terms
of Lorentz invariance violation by means of tensor-valued background fields fixed under
particle Lorentz transformations. These background fields can be interpreted as vacuum
expectation values that are generated by spontaneous symmetry breaking taking place in
a more fundamental theory. Studies in the SME have been pursued to look for LV effects
and to develop a precision programme that allows us to scrutinize the
limits of Lorentz symmetry in several physical interactions. In this sense, many investigations
were performed in the context of the SME fermion sector \cite{Fermion1,Fermion2},
\textit{CPT}-violating contributions~\cite{CPT}, the \textit{CPT}-odd electromagnetic
sector \cite{Adam1,Cherenkov1}, the \textit{CPT}-even electromagnetic sector
\cite{Cherenkov2,KM}, fermion-photon interactions \cite{KFint,Interac,Schreck1}, and
radiative corrections \cite{Radio}. Lorentz-violating theories are also connected to
higher-dimensional operators. In this sense, nonminimal extensions of the SME were
developed, both in the photon \cite{Kostelec1} and fermion sector \cite{Kostelec2}, composed
by \textit{CPT}-even and \textit{CPT}-odd higher-derivative operators. Other higher-dimensional
LV theories \cite{Myers1,Marat} were also proposed and examined. Nonminimal higher-dimensional
couplings that do not involve higher derivatives have been proposed and constrained, as well
\cite{NModd1,NM3,NMkost}.

Models of a higher-derivative electrodynamics have been investigated since the advent of
Podolsky's theory \cite{Podolsky1}, characterized by the Lorentz- and gauge-invariant term
$\partial_{\alpha} F^{\alpha\beta} \partial_{\lambda} F^{\lambda}{}_{\beta}$, one of the
simplest dimension-6 structures that can be built with the electromagnetic field. Podolsky's
Lagrangian is
\begin{equation}
\label{eq:lagrangian-podolsky}
\mathcal{L}=-\frac{1}{4}F_{\mu\nu}F^{\mu}{}^{\nu}+\frac{\theta^{2}}{2}%
\partial_{\alpha}F^{\alpha}{}^{\beta}\partial_{\lambda}%
F_{\phantom{\lambda}\beta}^{\lambda}-j_{\mu}A^{\mu}\,,
\end{equation}
where $F_{\mu\nu}=\partial_{\mu}A_{\nu}-\partial_{\nu}A_{\mu}$ is the field strength tensor
assigned to the vector field $A_{\mu}$ and $\theta$ is Podolsky's parameter having mass
dimension $-1$. The vector field is coupled to a conserved four-current $j_{\mu}$. One of
the most remarkable characteristics of Podolsky's electrodynamics is the generation of a
massive mode without the loss of gauge symmetry, being in this aspect different from Proca's
theory. The Podolsky propagator contains two poles, one corresponding to the massless photon
and the other one associated with the massive photon. At the classical level, the massive mode
of the model has the advantage of curing divergences connected to the pointlike self-energy,
but at the quantum level it is associated with the occurrence of ghosts \cite{Accioly1}. The suitable
gauge condition to address Podolsky's electrodynamics is not the usual Lorenz gauge, but a
modified gauge relation \cite{GalvaoP}, compatible with the existence of five degrees of
freedom (two related to a massless photon and three related to the massive mode). Other
aspects of quantum field theories in the presence of the Podolsky term, such as path integral
quantization and finite-temperature effects~\cite{Podolsky3,Podolsky4}, renormalization
\cite{Podolsky5}, as well as multipole expansion and classical solutions \cite{Podolsky6}
were also examined.

Another dimension-6 term, proposed in the late sixties, $F_{\mu\nu}\partial_{\alpha}\partial^{\alpha}F^{\mu\nu}$,
defines Lee-Wick electrodynamics \cite{LeeWick}, which leads to a finite self-energy for a
pointlike charge in (1+3) spacetime dimensions. Furthermore, it produces a
bilinear contribution to the Lagrangian similar to that of Podolsky's term but with opposite
sign. This \textquotedblleft wrong\textquotedblright\ sign yields energy instabilities at
the classical level, while it leads to negative-norm states in the Hilbert space at the
quantum level. Lee and Wick also proposed a mechanism to preserve unitarity, which removes all
states containing Lee-Wick photons from the Hilbert space. This theory regained attention
after the proposal of the Lee-Wick standard model \cite{LW}, based on a non-Abelian gauge
structure free of quadratic divergences. Such a model had a broad repercussion, with many
contributions in both the theoretical and phenomenological sense \cite{LW2}. In the Lee-Wick
scenario, studies of ghost states \cite{LWghost}, constructions endowed with higher derivatives
\cite{LWhd}, renormalization aspects \cite{LWren}, and finite-temperature investigations
\cite{LWft} have been reported, as well. The study of higher-derivative terms in quantum field
theories was also motivated by their role as ultraviolet regulators \cite{A.A}. Some works were
dedicated to investigating interactions between stationary sources in the context of Abelian
Lee-Wick models, with emphasis on sources distributed along parallel branes with arbitrary
dimension and the Dirac string in such a context \cite{Barone1}. Lee-Wick electrodynamics was
also studied for the case of the self-energy of pointlike charges in arbitrary dimensions,
exhibiting a finite ultraviolet result for $d=1$ and $d=3$ (spatial dimensions) \cite{Barone2},
for the case of perfectly conducting plates \cite{Barone3}, and in the evaluation of the
interaction between two pointlike charges \cite{Accioly2}. Recently, a LV higher-derivative and
dimension-6 term, $d_{\beta}d_{\sigma}\partial_{\alpha}F^{\alpha}{}^{\beta}\partial_{\lambda}
F^{\lambda\sigma}$  with an observer four-vector $d_{\mu}$, radiatively generated in
Ref.~\cite{Petrov}, was considered in the context of Maxwell's Lagrangian \cite{Barone4}. The
latter study focused on interactions between external sources with the modified electromagnetic
field, as performed in Refs.~\cite{Barone1,Barone2}.

For almost 20 years now, Lorentz-violating contributions of mass dimensions 3
and 4 have been investigated extensively from both a phenomenological and a theoretical point of
view. Many of the corresponding controlling coefficients are tightly constrained,
especially in the photon and lepton sector \cite{{Kostelecky:2008ts}}. Since a point in time not
long ago, a significant interest was aroused in field theories endowed with higher-order derivatives. In the \textit{CPT}-even
photon sector, the leading-order contributions in an expansion in terms of additional derivatives
are the dimension-6 ones. Hence, these are also the most prominent ones that could play a role
in nature, if higher-derivative Lorentz violation existed. Note that Lorentz-violating terms of mass dimension
6 are known to emerge from theories of noncommutative spacetimes, as the noncommutativity tensor
has a mass dimension of $-2$.

In the present work, we investigate basic features of a higher-derivative electrodynamics, in
which the physical fields are coupled to a \textit{CPT}-even and LV background by means of a
dimension-6 term. More specifically, we study Maxwell's theory modified by a LV operator of
mass dimension 6, which possesses two additional derivatives coupled to a \textit{CPT}-even
fixed background, $D_{\beta\alpha}$, in a structure of the form ${D_{\beta\alpha}}
\partial_{\sigma}F^{\sigma\beta}\partial _{\lambda} F^{\lambda\alpha}$. The latter is a kind of
anisotropic Podolsky term, i.e., it is a natural Lorentz-violating extension of
Podolsky's theory. Initially, the condition is discussed (nonzero trace) for which a LV
structure of this kind comprises the Podolsky term. It is interesting to note that in the
recent article \cite{Bonetti:2016vrq}, Lorentz violation is considered in a scenario with intact
supersymmetry. The Lorentz-violating background fields are assumed to be linked to a supersymmetric
multiplet and their effect on photon propagation is studied. After integrating out the contributions
from the photino, the effective Lagrangian given by their Eq.~(9) incorporates the type of modified
Podolsky term that we are studying. 

So far, not much is known about the properties of Lorentz-violating theories
including higher derivatives. It is a well-established fact, though, that higher-dimensional
operators lead to a rich plethora of new effects as well as additional issues. For example,
the existence of additional time derivatives may produce exotic modes that
cannot be considered as perturbations of the standard ones. These modes can lead to an indefinite
metric in Hilbert space, which is connected to the occurrence of states with negative norm.
The procedure developed by Lee and Wick \cite{LeeWick} makes it possible to deal with such modes
in a quantized theory such that a breakdown of unitarity is prevented. Before delving into these
possibly very profound problems of Lorentz-violating theories including higher derivatives, the classical
properties of these frameworks should be well understood first.

Describing classical aspects of LV theories with higher derivatives is the main motivation of the
current paper. Hence, we are interested in obtaining the Green's function of the field equations
and the dispersion relations as well as developing an understanding of classical causality. Using
a technique already employed in some previous LV models~\cite{Propagator}, which consists of finding
a closed projector algebra, and using the prescription $D_{\mu\nu}=({B_{\mu}}C_{\nu}+{B_{\nu}}C_{\mu})/2$
with two observer four-vectors $B_{\mu}$ and $C_{\mu}$, the propagator is derived and the
dispersion relations are determined from its poles. The goal of this work is to examine signal
propagation within a Podolsky electrodynamics modified by the term
${D_{\beta\alpha}}\partial_{\sigma}F^{\sigma\beta}\partial_{\lambda}F^{\lambda\alpha}$. Thus, the
modes described by the corresponding dispersion relations are
analyzed for several configurations of the fixed background and we search for sectors where the
energy is well-defined and causality is assured. Furthermore, we will perform a brief analysis of
unitarity of the theory for a vanishing Podolsky parameter. After doing so, we present a more
general dimension-6 higher-derivative Lagrangian that can be proposed in the presence of the
rank-2 background $D_{\beta\alpha}$. The latter also involves a kind of anisotropic Lee-Wick
term ${D^{\mu\nu}}\partial_{\sigma}F^{\sigma\lambda} \partial_{\mu}F_{\nu\lambda}$. The corresponding
propagator and the dispersion relations are derived again. Mode propagation is examined for several
configurations of the background, revealing that the dispersion relations of these
LV dimension-6 model may exhibit a physical behavior in the limit of large
momenta only. We finally show that the dimension-6 terms considered here are contained
in the nonminimal SME developed by Kosteleck\'{y} and Mewes~\cite{Kostelec1}. Throughout
the paper, natural units will be employed with $\hbar=c=1$.

\section{Maxwell electrodynamics modified by higher-derivative terms: some
possibilities}

As a first step, we propose a Maxwell electrodynamics modified by a
higher-derivative, \textit{CPT}-even term of mass dimension 6 including two
additional derivatives coupled to a fixed tensor $D_{\beta\alpha},$
that is,
\begin{equation}
{D_{\beta\alpha}}\partial_{\sigma}F^{\sigma}{}^{\beta}\partial_{\lambda}F^{\lambda}{}^{\alpha}\,, \label{TADLV1}%
\end{equation}
which represents a kind of anisotropic and generalized Podolsky term.
Without a restriction of generality, $D_{\beta\alpha}$ can be taken
to be symmetric, as its antisymmetric part does not contribute, anyhow. Hence, a
possible Lagrangian to be considered is
\begin{equation}
\mathcal{L}=-\frac{1}{4}F_{\mu\nu}F^{\mu}{}^{\nu}+\eta^{2}{D_{\beta\alpha}}
\partial_{\sigma}F^{\sigma}{}^{\beta}\partial_{\lambda}F^{\lambda}{}^{\alpha}\,, \label{LVSLAD1}%
\end{equation}
where the parameter $\eta$ has dimension of (mass)$^{\text{--1}}$. A property to
check is if the anisotropic piece (\ref{TADLV1}) contains a sector that is
equivalent to the Podolsky term of Eq.~(\ref{eq:lagrangian-podolsky}).
Such a term is generated from a nonvanishing trace of $D_{\mu\nu}$, what
can be shown quickly. If $D_{\mu\nu}$ contains nonzero diagonal components of the form
$D_{11}=D_{22}=D_{33}=-D_{00}$, the tensor involves a trace that
is given by
\begin{equation}
{D^{\lambda}_{\phantom{\lambda}\lambda}}=D_{00}-D_{ii}=4D_{00}\,.
\end{equation}
We define a new traceless tensor $\tilde{D}_{\beta\alpha}$ by subtracting the trace from the latter:
\begin{equation}
{\tilde{D}_{\beta\alpha}\equiv D_{\beta\alpha}-\frac{1}{4}g_{\beta\alpha}D^{\rho}_{\phantom{\rho}\rho}=D_{\beta\alpha}-g_{\beta\alpha}D_{00}\,,} \label{Dtraceless}%
\end{equation}
which fulfills $\tilde{D}^{\kappa}_{\phantom{\kappa}\kappa}=0$ and leads to
\begin{equation}
{\tilde{D}_{\beta\alpha}}\partial_{\sigma}F^{\sigma}{}^{\beta}\partial_{\lambda}F^{\lambda\alpha}%
={D_{\beta\alpha}}\partial_{\sigma}F^{\sigma}{}^{\beta}\partial_{\lambda}F^{\lambda\alpha}
-{D_{00}(\partial_{\sigma}F^{\sigma}{}^{\beta}\partial_{\lambda}F_{\phantom{\lambda}\beta}^{\lambda})}\,. \label{HDLVT4}%
\end{equation}
The second term on the right-hand side corresponds to a Podolsky term, i.e., such
a term appears in connection to the trace of $D_{\mu\nu}$, indeed.
In this sense, there are two possibilities that can be pursued, i.e., we can assess a
dimension-6 electrodynamics that either contains or does not contain the
Podolsky term. One option that exhibits, in principle, the same physical
content of Lagrangian~(\ref{LVSLAD1}) is to consider Podolsky's
electrodynamics modified by the traceless LV dimension-6 term
of Eq.~(\ref{Dtraceless}), that is,
\begin{equation}
\mathcal{L}=-\frac{1}{4}F_{\mu\nu}F^{\mu}{}^{\nu}+\frac{\theta^{2}}{2}%
\partial_{\alpha}F^{\alpha}{}^{\beta}\partial_{\lambda}%
F_{\phantom{\lambda}\beta}^{\lambda}+\eta^{2}{\tilde{D}_{\beta\alpha}}\partial_{\sigma}F^{\sigma}%
{}^{\beta}\partial_{\lambda}F^{\lambda}{}^{\alpha}\,.
\label{LVSLAD2}%
\end{equation}
A cleaner option including only the LV dimension-6 contribution in the
context of Maxwell's electrodynamics, is defined when the Podolsky sector is
zero, $D^{\kappa}_{\phantom{\kappa}\kappa}=0$, so that the Lagrangian to be
addressed is
\begin{equation}
\mathcal{L}=-\frac{1}{4}F_{\mu\nu}F^{\mu}{}^{\nu}+\eta^{2}{\tilde{D}_{\beta\alpha}}\partial_{\sigma
}F^{\sigma}{}^{\beta}\partial_{\lambda}F^{\lambda}{}^{\alpha}\,. \label{a2}%
\end{equation}
\begin{table}[t]
\centering
\begin{tabular}
[c]{lllll}%
\toprule & $C$ & $P$ & $T$ & \textit{CPT}\\
\colrule $D_{00}$ & $+$ & $+$ & $+$ & $\ +$\\
$D_{0i}$ & $+$ & $-$ & $-$ & $\ +$\\
$D_{ij}$ & $+$ & $+$ & $+$ & $\ +$\\
\botrule &  &  &  &
\end{tabular}
\caption{Transformation properties of the components of $D_{\mu\nu}$ under
\textit{C}, \textit{P}, \textit{T} and their combination.}%
\label{tab:transformation-properties}%
\end{table}
The components of the tensor $D_{\beta\alpha}$ can be classified in accordance
with the behavior under the discrete \textit{C}, \textit{P}, and \textit{T} operations. To do so, we
decompose the sum over the contracted indices in the term~(\ref{TADLV1})
into components of the electric and magnetic fields $\mathbf{E}$, $\mathbf{B}$:
\begin{align}
{D_{\beta\alpha}}\partial_{\sigma}F^{\sigma}{}^{\beta}\partial_{\lambda}F^{\lambda}{}^{\alpha
}  &  ={D_{00}}\partial_{a}E^{a}{}\partial_{b}E^{b}{}%
-2\epsilon^{bil}{D_{0i}}\partial_{a}E^{a}{}\partial_{b}B^{l}\nonumber\\
&  \phantom{{}={}}-2{D_{0i}}\partial_{a}E^{a}{}\partial_{0}E{}^{i}+{D_{ij}}\partial
_{0}E{}^{i}\partial_{0}E{}^{j}\nonumber\\
&  \phantom{{}={}}+2\epsilon^{bjl}{D_{ij}}\partial_{0}E{}^{i}\partial_{b}B^{l}
+\epsilon^{bjl}\epsilon^{aim}{D_{ij}}\partial_{a}B^{m}\partial_{b}B^{l}\,,
\label{Dcomp}%
\end{align}
remembering that $F^{a0}{}=E^{a},F^{b}{}^{j}=-\epsilon^{bjl}B^{l}$ with the three-dimensional Levi-Civita
symbol $\epsilon^{ijk}$. Under
charge conjugation (\textit{C}), the electric and magnetic fields behave according to
$\mathbf{E}\rightarrow-\mathbf{E}$ and $\mathbf{B}\rightarrow-\mathbf{B},$
while $\partial_{\mu}\rightarrow\partial_{\mu}.$ In this way, we notice that
the coefficients $D_{00},D_{0i},D_{ij}$ are \textit{C}-even. Under parity
(\textit{P}), $\mathbf{E}\rightarrow-\mathbf{E}$ and $\mathbf{B}%
\rightarrow\mathbf{B}$, $\partial_{a}\rightarrow-\partial_{a},\partial
_{0}\rightarrow\partial_{0},$ so that $D_{0i}$ is parity-odd, and $D_{00}$ and
$D_{ij}$ are parity-even. Under time reversal ($T),$ $\mathbf{B}%
\rightarrow-\mathbf{B}$, $\mathbf{E}\rightarrow\mathbf{E}$, with $\partial
_{a}\rightarrow\partial_{a},\partial_{0}\rightarrow-\partial_{0}.$ This
implies that $D_{0i}$ is $T$-odd, while $D_{00}$ and $D_{ij}$ are
\textit{T}-even. A summary of these properties can be found in
Tab.~\ref{tab:transformation-properties}.

\section{Propagator of the dimension-6 generalized Podolsky theory}

In this section, we consider the Maxwell Lagrangian modified by the Podolsky
and the dimension-6 anisotropic higher-derivative term (\ref{TADLV1}), given
as
\begin{equation}
\mathcal{L}=-\frac{1}{4}F_{\mu\nu}F^{\mu}{}^{\nu}
+\frac{\theta^2}{2}\partial_{\alpha}F^{\alpha}{}^{\beta}\partial_{\lambda}F_{\phantom{\lambda}\beta}^{\lambda}
+\eta^{2}D_{\beta\alpha}\partial_{\sigma}F^{\sigma}{}^{\beta}\partial_{\lambda}F^{\lambda\alpha}{}
+\frac{1}{2\xi}\left(\partial_{\mu}A^{\mu}\right)^2\,, \label{LPVL1a}%
\end{equation}
where the last contribution is introduced to fix the gauge.\footnote{We mentioned in the
introduction that gauge conditions used in Maxwell's electrodynamics may cause problems
in Podolsky's extension~\cite{GalvaoP}. For example, as the vector field component $A^0$ is nondynamical,
Lorenz gauge requires that the solutions of the field equations be transverse, which is
not the case for massive modes. It is paramount to consider alternative gauge conditions
when quantizing the theory. However, we neither obtain the solutions of the equations of
motion nor do we quantize the framework under investigation. The focus is on studying the
modified dispersion relations, which is a classical analysis. Also, the gauge condition
will not modify the dispersion relations. Therefore, to reduce technical complications in
computing the propagator, we will still employ the usual set of gauges used in Maxwell's
electrodynamics.} The parameters $\theta$,
$\eta$ have dimension of (mass)$^{\text{--1}}$ and $D_{\beta\alpha}$ is the
dimensionless \textit{CPT}-even tensor introduced before. The coefficients $\theta^2$ and $\eta^2$
are here considered as positive in analogy with Podolsky's theory, where $\theta^2>0$ is a necessary condition for obtaining a physical dispersion relation. In a broad context, there exists the possibility of considering $\theta^2$ and $\eta^2$ as negative. These choices have the potential for altering dispersion relations and other
physical properties of the theory such as unitarity. However, an investigation of this sector is beyond the scope of the paper, which is why we
will assume that both $\theta^2\geq 0$ and $\eta^2\geq 0$. The Lagrangian~(\ref{LPVL1a}) can be
written in the bilinear form,
\begin{equation}
\label{eq:lagrange-density-bilinear}
\mathcal{L}=\frac{1}{2}A^{\nu}O_{\mu\nu}A^{\mu}\,,
\end{equation}
where $O_{\mu\nu}$ is%
\begin{align}
O_{\mu\nu} &  ={(1+\theta^2\square)\square\Theta_{\mu\nu}}-\frac{1}{\xi}\square\Omega_{\mu\nu}\nonumber\\
&\phantom{{}={}}+2\eta^2[{D_{\nu\mu}}\square^2
-{D_{\nu\alpha}}\square\partial_{\mu}\partial^{\alpha}
-{D_{\sigma\mu}}\square\partial_{\nu}\partial^{\sigma}
+{D_{\sigma\alpha}}\partial_{\nu}\partial_{\mu}\partial^{\sigma}\partial^{\alpha}]\,. \label{OLV1}%
\end{align}
\begin{table}[b]
\centering
\[%
\begin{array}
[c]{cccccc}%
\toprule & \Theta_{\phantom{\nu}\alpha}^{\nu} & \Omega_{\phantom{\nu}\alpha
}^{\nu} & B_{\alpha}\partial^{\nu} & B_{\alpha}C^{\nu} & C^{\nu}%
\partial_{\alpha}\\
\colrule \Theta_{\mu\nu} & \Theta_{\mu\alpha} & 0 & 0 & B_{\alpha}C_{\mu}
-\frac{\rho}{\square}B_{\alpha}\partial_{\mu} & C_{\mu}\partial_{\alpha}%
-\rho\Omega_{\mu\alpha}\\
\colrule \Omega_{\mu\nu} & 0 & \Omega_{\mu\alpha} & B_{\alpha}\partial_{\mu} &
\frac{\rho}{\square}B_{\alpha}\partial_{\mu} & \rho\Omega_{\mu\alpha}\\
\colrule B_{\nu}\partial_{\mu} & B_{\alpha}\partial_{\mu}-\kappa\Omega
_{\mu\alpha} & \kappa\Omega_{\mu\alpha} & \kappa B_{\alpha}\partial_{\mu} &
(B\cdot C)  B_{\alpha}\partial_{\mu} & (B\cdot
C)\square\Omega_{\mu\alpha}\\
\colrule B_{\nu}C_{\mu} & B_{\alpha}C_{\mu}-\frac{\kappa}{\square}C_{\mu
}\partial_{\alpha} & \frac{\kappa}{\square}C_{\mu}\partial_{\alpha} & \kappa
B_{\alpha}C_{\mu} & (B\cdot C)B_{\alpha}C_{\mu} & (
B\cdot C)  C_{\mu}\partial_{\alpha}\\
\colrule C_{\mu}\partial_{\nu} & 0 & C_{\mu}\partial_{\alpha} & \square
B_{\alpha}C_{\mu} & \rho B_{\alpha}C_{\mu} & \rho C_{\mu}\partial_{\alpha}\\
\colrule B_{\mu}\partial_{\nu} & 0 & B_{\mu}\partial_{\alpha} & \square
B_{\mu}B_{\alpha} & \rho B_{\mu}B_{\alpha} & \rho B_{\mu}\partial_{\alpha}\\
\colrule B_{\mu}C_{\nu} & B_{\mu}C_{\alpha}-\frac{\rho}{\square}B_{\mu
}\partial_{\alpha} & \frac{\rho}{\square}B_{\mu}\partial_{\alpha} & \rho
B_{\mu}B_{\alpha} & C^2B_{\mu}B_{\alpha} & C^2B_{\mu}\partial_{\alpha}\\
\colrule C_{\nu}\partial_{\mu} & C_{\alpha}\partial_{\mu}-\rho\Omega
_{\mu\alpha} & \rho\Omega_{\mu\alpha} & \rho B_{\alpha}\partial_{\mu} &
C^{2}B_{\alpha}\partial_{\mu} & C^2\square\Omega_{\mu\alpha}\\
\colrule B_{\mu}B_{\nu} & B_{\mu}B_{\alpha}-\frac{\kappa}{\square}B_{\mu
}\partial_{\alpha} & \frac{\kappa}{\square}B_{\mu}\partial_{\alpha} & \kappa
B_{\mu}B_{\alpha} & (B\cdot C)  B_{\mu}B_{\alpha} & (
B\cdot C)  B_{\mu}\partial_{\alpha}\\
\colrule C_{\mu}C_{\nu} & C_{\mu}C_{\alpha}-\frac{\rho}{\square}C_{\mu
}\partial_{\alpha} & \frac{\rho}{\square}C_{\mu}\partial_{\alpha} & \rho
B_{\alpha}C_{\mu} & C^2B_{\alpha}C_{\mu} & C^2C_{\mu}\partial_{\alpha}\\
\botrule
\end{array} \]
\caption{Closed algebra of tensor projectors (part 1).}%
\label{tab:closed-algebra-1}%
\end{table}
Here, we have used the longitudinal and transverse projectors,
\begin{equation}
\label{eq:projectors}
\Omega_{\beta\lambda}=\frac{\partial_{\beta}\partial_{\lambda}}{\square}\,,\quad
\Theta_{\beta\lambda}=g_{\beta\lambda}-\Omega_{\beta\lambda}\,,
\end{equation}
respectively, where $g_{\mu\nu}$ is the Minkowski metric with signature $(+,-,-,-)$.
To derive the propagator, we propose the following
parameterization:%
\begin{equation}
D_{\mu\nu}=\frac{1}{2}({B_{\mu}}C_{\nu}+{B_{\nu}}C_{\mu})\,,
\label{Presc1}%
\end{equation}
where $C_{\mu},B_{\nu}$ are two independent observer four-vectors giving rise to preferred spacetime
directions. This parameterization is nearly general and describes most of the configurations of the
symmetric $D_{\beta\alpha}$ tensor. It is used for technical reasons, mainly connected to the construction of the
propagator of this theory. Furthermore, it allows us to
classify different sectors of the theory by geometric properties related to the two vectors, e.g.,
orthogonality of their spatial parts. With the latter choice, the operator of Eq.~(\ref{OLV1}) becomes
\begin{align}
O_{\mu\nu}  &  =(1+\theta^2\square)\square\Theta_{\mu\nu}+\left(2\eta^2{\kappa\rho}-\frac{1}{\xi}\right)\square\Omega_{\mu\nu}
+\eta^2{(B_{\mu}C_{\nu}+B_{\nu}C_{\mu})}\square^2 \nonumber\\
&\phantom{{}={}}-\eta^2\square({C_{\nu}}\kappa \partial_{\mu}
+{C_{\mu}}\kappa\partial_{\nu}+{B_{\nu}}\rho\partial_{\mu}
+{B_{\mu}}\rho\partial_{\nu})\,,
\label{OLV2}%
\end{align}
with%
\begin{equation}
\kappa=B_{\alpha}\partial^{\alpha}\,,\quad\rho=C_{\alpha}\partial^{\alpha}\,.
\label{karo}%
\end{equation}
\begin{table}[ptb]
\centering
\[%
\begin{array}
[c]{cccccc}%
\toprule & B^{\nu}\partial_{\alpha} & B^{\nu}C_{\alpha} & C_{\alpha}%
\partial^{\nu} & B^{\nu}B_{\alpha} & C^{\nu}C_{\alpha}\\
\colrule \Theta_{\mu\nu} & B_{\mu}\partial_{\alpha}-\kappa\Omega_{\mu\alpha} &
B_{\mu}C_{\alpha}-\frac{\kappa}{\square}C_{\alpha}\partial_{\mu} & 0 & B_{\mu
}B_{\alpha}-\frac{\kappa}{\square}B_{\alpha}\partial_{\mu} & C_{\mu}C_{\alpha
}-\frac{\rho}{\square}C_{\alpha}\partial_{\mu}\\
\colrule \Omega_{\mu\nu} & \kappa\Omega_{\mu\alpha} & \frac{\kappa}{\square
}C_{\alpha}\partial_{\mu} & C_{\alpha}\partial_{\mu} & \frac{\kappa}{\square
}B_{\alpha}\partial_{\mu} & \frac{\rho}{\square}C_{\alpha}\partial_{\mu}\\
\colrule B_{\nu}\partial_{\mu} & \square B^{2}\Omega_{\mu\alpha} &
B^2C_{\alpha}\partial_{\mu} & \kappa C_{\alpha}\partial_{\mu} &
B^2B_{\alpha}\partial_{\mu} & (B\cdot C)  C_{\alpha}%
\partial_{\mu}\\
\colrule B_{\nu}C_{\mu} & B^2C_{\mu}\partial_{\alpha} & B^2C_{\mu
}C_{\alpha} & \kappa C_{\mu}C_{\alpha} & B^2B_{\alpha}C_{\mu} & (
B\cdot C)  C_{\mu}C_{\alpha}\\
\colrule C_{\mu}\partial_{\nu} & \kappa C_{\mu}\partial_{\alpha} & \kappa
C_{\mu}C_{\alpha} & \square C_{\mu}C_{\alpha} & \kappa B_{\alpha}C_{\mu} &
\rho C_{\mu}C_{\alpha}\\
\colrule B_{\mu}\partial_{\nu} & \kappa B_{\mu}\partial_{\alpha} & \kappa
B_{\mu}C_{\alpha} & \square B_{\mu}C_{\alpha} & \kappa B_{\mu}B_{\alpha} &
\rho B_{\mu}C_{\alpha}\\
\colrule B_{\mu}C_{\nu} & (B\cdot C)B_{\mu}\partial_{\alpha} &
(B\cdot C)B_{\mu}C_{\alpha} & \rho B_{\mu}C_{\alpha} & (
B\cdot C) B_{\mu}B_{\alpha} & C^2B_{\mu}C_{\alpha}\\
\colrule C_{\nu}\partial_{\mu} & \square(B\cdot C)  \Omega
_{\mu\alpha} & (B\cdot C)C_{\alpha}\partial_{\mu} & \rho
C_{\alpha}\partial_{\mu} & (B\cdot C)B_{\alpha}\partial_{\mu} &
C^2C_{\alpha}\partial_{\mu}\\
\colrule B_{\mu}B_{\nu} & B^2B_{\mu}\partial_{\alpha} & B^2B_{\mu}C_{\alpha} &
\kappa B_{\mu}C_{\alpha} & B^2B_{\mu}B_{\alpha} & (B\cdot
C)B_{\mu}C_{\alpha}\\
\colrule C_{\mu}C_{\nu} & (B\cdot C)C_{\mu}\partial_{\alpha} &
(B\cdot C)C_{\mu}C_{\alpha} & \rho C_{\mu}C_{\alpha} &
(B\cdot C)B_{\alpha}C_{\mu} & C^2C_{\mu}C_{\alpha}\\
\botrule
\end{array} \]
\caption{Closed algebra of tensor projectors (part 2).}%
\label{tab:closed-algebra-2}%
\end{table}
To derive the propagator, we need to invert the operator
$O_{\mu\nu},$ composed of the projectors $\Theta_{\mu\nu}$, $\Omega_{\mu\nu}$,
$B_{\nu}\partial_{\mu}$, $B_{\mu}\partial_{\nu}$, $C_{\mu}B_{\nu}$, $C_{\nu}B_{\mu}$,
$C_{\mu}\partial_{\nu}$, $C_{\nu}\partial_{\mu}$. In this sense, the \textit{Ansatz}
\begin{align}
\Delta_{\phantom{\nu}\alpha}^{\nu}  &  =a\Theta_{\phantom{\nu}\alpha}^{\nu}
+b\Omega_{\phantom{\nu}\alpha}^{\nu}+cB_{\alpha}\partial^{\nu}
+dC^{\nu}B_{\alpha}+eC^{\nu}\partial_{\alpha}\nonumber\\
&\phantom{{}={}}+fB^{\nu}\partial_{\alpha}+gC_{\alpha}B^{\nu}+hC_{\alpha}\partial^{\nu}
+iB^{\nu}B_{\alpha}+jC^{\nu}C_{\alpha}\,, \label{PropLV3}%
\end{align}
for the Green's function $\Delta_{\nu\alpha}=\Delta_{\nu\alpha}(x-y)$ is proposed obeying the condition, $O_{\mu\nu}\Delta_{\text{ }}^{\nu\alpha}=\delta
_{\mu}^{\phantom{\mu}\alpha},$ or%
\begin{equation}
O_{\mu\nu}\Delta_{\phantom{\nu}\alpha}^{\nu}=\Theta_{\mu\alpha}+\Omega_{\mu\alpha}\,. \label{Prop1a}%
\end{equation}
The tensor projectors contained in Eq.~(\ref{PropLV3}) fulfill a closed
algebra, as shown in Tabs.~\ref{tab:closed-algebra-1}, \ref{tab:closed-algebra-2}.
By inserting Eq.~(\ref{PropLV3}) into Eq.~(\ref{Prop1a}), we obtain a system of ten
equations for the ten coefficients to be determined, whose solution is given
by
\begin{subequations}
\begin{align}
a &  =\frac{1}{\square(1+\theta^2\square)}\,,\quad
c=f=\frac{\eta^4({C^2}\square-\rho^2){\kappa}-\eta^2\rho\Pi}{\Gamma\square(1+\theta^2\square)}\,,\displaybreak[0]\\[2ex]
b &  =-\frac{\xi}{\square}+\frac{\eta^2\left[-\eta^2\rho^2({B^2}\square-\kappa^2)-\eta^2\kappa^2({C^2}\square-\rho^2)
+2{\kappa\rho}\Pi\right]}{\Gamma\square(1+\theta^2\square)}\,,\displaybreak[0]\\[2ex]
d &  =g=\frac{\eta^2\Pi}{(1+\theta^2\square)\Gamma}\,,\quad e=h=\eta^2\frac{[\eta^2\rho({B^2}\square-\kappa^2)
-{\kappa}\Pi]}{\Gamma\square(1+\theta^2\square)}\,,\displaybreak[0]\\[2ex]
i &  =-\frac{\eta^4({C^2}\square-\rho^2)}{\Gamma(1+\theta^2\square)}\,,\quad
j=-\frac{\eta^4({B^2}\square-\kappa^2)}{\Gamma(1+\theta^2\square)}\,,
\end{align}
where%
\end{subequations}
\begin{subequations}
\begin{align}
\Gamma &  =\eta^4({B^2}\square
-\kappa^2)({C^2}\square-\rho^2)-\Pi^2,\displaybreak[0]\\[2ex]
\Pi &  =1+\theta^2\square+\eta^2{(B\cdot C)}\square-\eta^2{\kappa\rho}\,.
\end{align}
In momentum space, the propagator is%
\end{subequations}
\begin{align}
\Delta_{\nu\alpha}&=-\frac{\mathrm{i}}{p^2(1-\theta^2p^2)\Gamma(p)}%
\Big\{\Gamma(p)\Theta_{\nu\alpha}+\left[b^{\prime}-\xi(1-\theta^2p^2)\Gamma(p)\right]\Omega_{\nu\alpha}\nonumber\\
&\phantom{{}={}-p^2(1-\theta^2p^2)\Gamma(p)\big\{}-\mathrm{i}%
F(p)(B_{\nu}p_{\alpha}+B_{\alpha}p_{\nu})-2\eta^{2}{D_{\nu\alpha}}p^{2}%
\Pi(p)-\mathrm{i}H(p)(C_{\nu}p_{\alpha}+C_{\alpha}p_{\nu})\nonumber\\[1ex]
&\phantom{{}={}-p^2(1-\theta^2p^2)\Gamma(p)\big\{}\left.
+\,\eta^4p^2{B_{\nu}B_{\alpha}}\left[(C\cdot p)^2-{C^2}p^2\right]
+\eta^4p^2{C_{\nu}C_{\alpha}}\left[(B\cdot p)^2%
-{B^2}p^2\right]\right\}\,, \label{PropLV5}%
\end{align}
where $p_{\mu}$ is the four-momentum and
\begin{subequations}
\label{eq:definitions-bprime-f-h}%
\begin{align}
b^{\prime}&=\eta^2\left\{\eta^2(C\cdot p)^2\left[(B\cdot p)^2-{B^2}p^2\right]\right.
\displaybreak[0]\label{blinha}\\
&\phantom{{}={}\eta^2\big\{}\left.+\,\eta^{2}(B\cdot p)^2\left[(C\cdot p)^2-{C^2}p^2\right]
-2{(B\cdot p)}(C\cdot p)\Pi(p)\right\}\,,\displaybreak[0]\nonumber\\[2ex]
F(p)&=\mathrm{i}\eta^{2}{(C\cdot p)}\Pi(p)-\mathrm{i}\eta^{4}(B\cdot p)\left[(C\cdot p)^2
-{C^2}p^2\right] \,,\label{Fp}\displaybreak[0]\\[2ex]
H(p)&=\mathrm{i}\eta^2(B\cdot p)\Pi(p)-\mathrm{i}\eta^4(C\cdot p)\left[(B\cdot p)^2-{B^2}p^2\right]\,, \label{Hp}%
\end{align}
with
\end{subequations}
\begin{subequations}
\label{eq:definitions-gamma-pi}%
\begin{align}
\Gamma(p)& =\eta^4{\left[(B\cdot p)^2-B^2p^2\right]}\left[(C\cdot p)^2-{C^2}p^2\right]
-\Pi^2(p)\,,\displaybreak[0]\label{lambp}\\[2ex]
\Pi(p)&=1-\theta^2p^2-\eta^2{(B\cdot C)}p^2+\eta^2{(B\cdot p)}(C\cdot p)\,.
\end{align}
Note that $F$ and $H$ have dimensions of (mass)$^{-1}$, while $\Gamma$ and
$\Pi$ are dimensionless. In the absence of the LV term,
$\eta=0$, and the propagator (\ref{PropLV5}) takes the form,
\end{subequations}
\begin{equation}
\Delta_{\nu\alpha}=-\mathrm{i}\left(\frac{\Theta_{\nu\alpha}}{p^2(1-\theta^2p^2)}-\frac{\xi}{p^2}\Omega_{\nu\alpha}\right)\,,
\label{PPodolsky}
\end{equation}
recovering Podolsky's propagator, as expected. Setting $\theta^{2}=0$ in the
propagator (\ref{PropLV5}), the result is
\begin{align}
\Delta_{\nu\alpha}&=-\frac{\mathrm{i}}{p^2\Gamma(p)}\Big\{\Gamma(p)\Theta_{\nu\alpha}+(b^{\prime}-\xi\Gamma)\Omega_{\nu\alpha}\nonumber\\
&\phantom{{}={}-p^{2}\Gamma(p)}\,-\mathrm{i}F(p)(B_{\nu}p_{\alpha}+B_{\alpha}p_{\nu})-2\eta^2{D_{\nu\alpha}}p^{2}\Pi(p)
-\mathrm{i}H(p)(C_{\nu}p_{\alpha}+C_{\alpha}p_{\nu})\nonumber\\[1ex]
&\phantom{{}={}-p^2\Gamma(p)}\left.\,+\,\eta^4{B_{\nu}B_{\alpha}}\left[(C\cdot p)^2-{C^2}p^2\right]{p^2}
+\eta^4{C_{\nu}C_{\alpha}}\left[(B\cdot p)^2-{B^2}p^2\right]{p^2}\right\}\,, \label{PropLV6}%
\end{align}
where $b^{\prime},F\left(  p\right)  ,H\left(  p\right)  ,\Gamma\left(
p\right)  $ are given by the same expressions of
Eqs.~(\ref{eq:definitions-bprime-f-h}), (\ref{eq:definitions-gamma-pi}),
with
\begin{equation}
\Pi_{\theta=0}(p)=1-\eta^{2}{(B\cdot C)}p^{2}  +\eta^{2}{(B\cdot p)}(C\cdot p)\,. \label{P3}%
\end{equation}
In this situation there are still two poles, $p^{2}=0,$ $\Gamma\left(
p\right)  =0,$ associated with the Maxwell modes and those related to the LV
higher-derivative term. Thus, in principle, the Lagrangian (\ref{PropLV6}) has
degrees of freedom linked to both modes, yielding a counting of modes analogue
to that in Podolsky's electrodynamics (see Ref.~\cite{GalvaoP}).

\subsection{Dispersion relations}

The dispersion equations of the modified electrodynamics defined by Eq.~(\ref{LPVL1a})
can be red off the poles of the propagator~(\ref{PropLV5}) in momentum space,
\begin{subequations}
\label{eq:n3}
\begin{align}
p^2(1-\theta^2p^2)&=0\,,\label{n3a}\\[2ex]
\Gamma(p)&=0\,. \label{n3b}%
\end{align}
\end{subequations}
Since these poles also appear in terms that are not connected to the
gauge fixing parameter $\xi$, they must be physical.
As before, the dispersion equation $p^{2}=0$ represents the well-known Maxwell modes,
while $1-\theta^{2}p^{2}=0$ stands for the Podolsky modes. Both are not
modified by the higher-derivative term, whose effect is fully encoded in the
dispersion equation~(\ref{n3b}). In the absence of the Podolsky term, $\theta^{2}=0,$ the
modified Eq.~(\ref{n3b}) yields:%
\begin{equation}
\eta^{4}{\left[(B\cdot p)^2-B^2p^2\right]\left[(C\cdot p)^2-C^2p^2\right]}-\left[  1-\eta^2%
{(B\cdot C)}p^2+\eta^2{(B\cdot p)}(C\cdot p)\right]^2=0\,. \label{n4}%
\end{equation}
With the help of \textit{FORM}\footnote{\textit{FORM} is a programming language that allows for symbolic manipulations
of mathematical expressions to be performed. It is widely used for evaluating lengthy algebraic expressions that occur in computations
of quantum corrections in high-energy physics.} \cite{Vermaseren:2000nd},
the dispersion equations above can be generalized to an arbitrary choice of $D_{\mu\nu}$. For a general
choice of this tensor, the dispersion equation $p^2=0$ remains, but $1-\theta^2p^2=0$ and Eq.~(\ref{n4}) merge into a single equation.
Although, in principle, there are many possibilities of forming observer Lorentz scalars from a general two-tensor and the momentum
four-vector, the latter result collapses when taking into account that $D_{\mu\nu}$ is symmetric. The equation can then be conveniently written as
follows:
\begin{align}
\label{4n}
0&=-\,(1-\theta^2p^2)^3-2(1-\theta^2p^2)^2\eta^2(D_{\mu\nu}p^{\mu}p^{\nu}-p^2D^{\mu}_{\phantom{\mu}\mu}) \notag \\
&\phantom{{}={}}+2(1-\theta^2p^2)p^2\eta^4\left[p^2D_{\mu\nu}D^{\mu\nu}-2D_{\mu\nu}D^{\mu\varrho}p^{\nu}p_{\varrho}+D^{\mu}_{\phantom{\mu}\mu}\left(2D_{\nu\varrho}p^{\nu}p^{\varrho}-p^2D^{\nu}_{\phantom{\nu}\nu}\right)\right] \notag \\
&\phantom{{}={}}+\frac{4}{3}p^4\eta^6\left[3D_{\mu\nu}D^{\mu\nu}\left(D_{\varrho\sigma}p^{\varrho}p^{\sigma}-p^2D^{\varrho}_{\phantom{\varrho}\varrho}\right)+(D^{\mu}_{\phantom{\mu}\mu})^2\left(p^2D^{\nu}_{\phantom{\nu}\nu}-3D_{\nu\varrho}p^{\nu}p^{\varrho}\right)\right. \notag \\
&\phantom{{}={}+\frac{4}{3}p^4\eta^6\Big[}\left.+\,2p^2D^{\mu\nu}D_{\mu}^{\phantom{\mu}\varrho}D_{\nu\varrho}+6\left(D_{\mu\nu}D^{\mu\varrho}p^{\nu}p_{\varrho}D^{\sigma}_{\phantom{\sigma}\sigma}-D_{\mu\nu}D^{\mu\varrho}D^{\nu\sigma}p_{\varrho}p_{\sigma}\right)\right]\,.
\end{align}
Note that it is not possible to factor out $1-\theta^2p^2$, as there is a contribution proportional to $\eta^6$ that does not contain the latter term.
This contribution vanishes when $D_{\mu\nu}$ is decomposed into two four-vectors according to Eq.~(\ref{Presc1}), whereupon $(1-\theta^2p^2)\Gamma(p)=0$
is reproduced.

\subsection{Analysis of some sectors of the theory}
\label{sec:analysis-sectors}

In this section, we analyze some sectors of Eqs.~(\ref{eq:n3}) -- (\ref{4n}) to search for LV configurations that exhibit a consistent
physical behavior. We are especially interested in causality. The behavior of the group and front velocity~\cite{Brillouin:1960}
allows for conclusions to be drawn on it where
\begin{equation}
\mathbf{u}_{\mathrm{gr}}\equiv\frac{\partial p_{0}}{\partial\mathbf{p}%
}\,,\quad u_{\mathrm{fr}}\equiv\lim_{|\mathbf{p}|\mapsto\infty}\frac{p_{0}%
}{|\mathbf{p}|}\,.
\end{equation}
We use a notion of classical causality that requires $u_{\mathrm{gr}}\equiv|\mathbf{u}_{\mathrm{gr}}|\leq1$ and $u_{\mathrm{fr}}\leq1$.
Dispersion laws that do not describe standard photons for vanishing Lorentz-violation will
be referred to as ``exotic,'' which includes Podolsky-type dispersion relations, but not necessarily unphysical ones.
Dispersion relations that exhibit divergent group/front velocities or velocities greater than 1 will be called ``spurious.''
For all the investigated background configurations, there are three distinct poles, including $p^{2}=0$, which describes the usual
photon.\footnote{This fact will be demonstrated in Sec.~\ref{sec:unitarity-analysis} explicitly.} The first and the second of the
isotropic cases examined below cannot be parameterized with two four-vectors as proposed in Eq.~(\ref{Presc1}). Therefore, they must be
studied by using the more general Eq.~(\ref{4n}).

\subsubsection{Isotropic trace sector}
\label{sec:isotropic-case-sector}

As an initial cross check of Eq.~(\ref{4n}), we study the case of a diagonal tensor
$D_{\beta\alpha}$ with $D_{00}=-D_{11}= -D_{22}=-D_{33}$, $D_{ij}=0$ for $i\neq j$, i.e.,
$D_{\beta\alpha}= D_{00}g_{\beta\alpha}$, which gives $D^{\kappa}_{\phantom{\kappa}\kappa}=4D_{00}$. It is a pure
trace configuration that is Lorentz-invariant, though. The dispersion equation for this configuration of $D_{\mu\nu}$ reads:
\begin{equation}
\left[1-\Theta^2(p_0^2-\mathbf{p}^2)\right]^3=0\,,\quad\Theta^2=\theta^2+2\eta^2D_{00}\,,
\end{equation}
from which we deduce
\begin{equation}
\label{n5}%
p_{0}=\sqrt{\mathbf{p}^{2}+\frac{1}{\Theta^2}}\,.
\end{equation}
This result is Podolsky's dispersion relation, obviously with a redefined Podolsky parameter $\Theta$ that involves
the standard Podolsky parameter $\theta$ and the nonvanishing controlling coefficient. If a Lorentz-violating background field
$D_{\mu\nu}$ existed in nature, its trace would mimic the Podolsky term. Finding such a contribution experimentally, could
even hint towards the existence of a Lorentz-violating background. The dispersion relation found is causal and compatible with a
well-behaved propagation of signals as long as $\Theta^2>0$, i.e., for $D_{00}>-\theta^2/(2\eta^2)$.

\subsubsection{Complete isotropic sector}

The traceless isotropic sector can be expressed in the form $D_{\mu\nu}=-D_{00}\times\mathrm{diag}(3,1,1,1)_{\mu\nu}$
where the global minus sign is chosen to be consistent with the definitions in Sec.~\ref{sec:isotropic-case-sector}. This choice of
the background tensor is traceless and can be covariantly expressed in terms of the Minkowski metric tensor and the purely timelike
preferred direction $\lambda^{\mu}=(1,0,0,0)^{\mu}$ as follows:
\begin{equation}
D_{\mu\nu}=D_{00}\left(g_{\mu\nu}-4\lambda_{\mu}\lambda_{\nu}\right)\,.
\end{equation}
Hence, the isotropic case under consideration contains the Lorentz-invariant part considered in Sec.~\ref{sec:isotropic-case-sector} and
a Lorentz-violating contribution dependent on the preferred direction $\lambda_{\mu}$, which is similar to that of
Sec.~\ref{sec:timelike-isotropic-sector}. This particular framework exhibits two distinct dispersion relations that can be obtained
from Eq.~(\ref{4n}):
\begin{subequations}
\begin{align}
p_0^{(1)}&=\sqrt{\mathbf{p}^2+\frac{1}{\Theta^2}}\,,\quad \Theta^2=\theta^2+2\eta^2D_{00}\,, \\[2ex]
p_0^{(2)}&=\sqrt{\Lambda\mathbf{p}^2+\frac{1}{\Theta^2}}\,,\quad \Lambda=1-\frac{8\eta^2D_{00}}{\Theta^2}\,.
\end{align}
\end{subequations}
The first dispersion law is a perturbation of the usual Podolsky dispersion relation, which corresponds to Eq.~(\ref{n5}) found before.
We already know that this mode has a well-defined energy and is causal for $D_{00}\geq -\theta^2/(2\eta^2)$. The second dispersion law
is a perturbation, as well, but it is at bit more involved, as the momentum-dependent terms are also modified. The front velocity of the
latter is given by $u_{\mathrm{fr}}^{(2)}=\sqrt{\Lambda}$. Thus,
the front velocity is real for $D_{00}\leq \theta^2/(6\eta^2)$ and it is $\leq 1$ as long as $D_{00}>0$. The modulus of the
group velocity can be cast into the form $u_{\mathrm{gr}}^{(2)}=\Lambda|\mathbf{p}|/p_0^{(2)}$. This function monotonically increases
from 0 to a constant value given by
\begin{equation}
\lim_{|\mathbf{p}|\mapsto\infty} u_{\mathrm{gr}}^{(2)}=u_{\mathrm{fr}}^{(2)}=\sqrt{\frac{\theta^2-6\eta^2D_{00}}{\theta^2+2\eta^2D_{00}}}\,.
\end{equation}
The latter quantity is $\leq 1$ for $D_{00}>0$, i.e., the framework is causal and well-behaved for a controlling coefficient in the range
$D_{00}\in [0,\theta^2/(6\eta^2)]$.

\subsubsection{Timelike isotropic sector}
\label{sec:timelike-isotropic-sector}

The simplest isotropic LV configuration is based on the decomposition of Eq.~(\ref{Presc1}) with the purely timelike directions
$B_{\mu}=(B_0,\mathbf{0})_{\mu}$, $C_{\mu}=(C_0,\mathbf{0})_{\mu}$. This sector corresponds to $D_{00}=C_0B_0$ and $D_{0i}=D_{i0}=D_{ij}=0$.
For the latter, Eq.~(\ref{eq:n3}) can be solved for $p_0$ providing
\begin{subequations}
\begin{align}
p_0^{(1)}&=\sqrt{\mathbf{p}^2+\frac{1}{\theta^2}}\,, \\[2ex]
p_0^{(2)}&=\sqrt{\Psi\mathbf{p}^2+\frac{1}{\theta^2}}\,,\quad \Psi=1+\frac{2\eta^2D_{00}}{\theta^2}\,,
\end{align}
\end{subequations}
where $p_0^{(1)}$ is the original Podolsky dispersion relation and $p_0^{(2)}$ is a perturbation of the latter. The first is known to
describe a proper propagation of signals. The front velocity of
the second is $u_{\mathrm{fr}}^{(2)}=\sqrt{\Psi}$, which requires $D_{00}$ to be nonpositive such that $u_{\mathrm{fr}}^{(2)}\leq 1$.
Furthermore, the modulus of the group velocity amounts to $u_{\mathrm{gr}}^{(2)}=\Psi|\mathbf{p}|/p_0^{(2)}$. It rises
monotonically from 0 to a constant value that corresponds to the front velocity of this mode:
\begin{equation}
\lim_{|\mathbf{p}|\mapsto\infty} u_{\mathrm{gr}}^{(2)}=u_{\mathrm{fr}}^{(2)}\,.
\end{equation}
Therefore, $D_{00}\leq 0$ has to hold to grant causal signal propagation.

\subsubsection{Parity-even anisotropic sector (with $\theta=0)$}
\label{sec:parity-even-anisotropic}

For the configuration described by the two purely spacelike directions $B_{\mu}=(0,\mathbf{B})_{\mu}$ and $C_{\mu}=(0,\mathbf{C})_{\mu}$, the
components of the two-tensor background field are $D_{00}=0$, $D_{0i}=D_{i0}=0$, and
$D_{ij}=({B_i}C_j+{B_j}C_i)/2$. Furthermore, the preferred directions obey
$B^2=-\mathbf{B}^2$, $C^2=-\mathbf{C}^2$, $B\cdot p=-\mathbf{B\cdot
p}$, and $C\cdot p=-\mathbf{C\cdot p}$. Inserting this information into
Eq.~(\ref{n4}) delivers:
\begin{align}
0&=\eta^{4}\left[{\mathbf{B}^2}\mathbf{C}^2p^4+{\mathbf{B}^2}(\mathbf{C\cdot p})^2p^2
+\mathbf{C}^2(\mathbf{B\cdot p})^2p^2+{(\mathbf{B\cdot p})^2}(\mathbf{C\cdot p})^2\right] \nonumber\\
&\phantom{{}={}}-\left[1+\eta^2{(\mathbf{B\cdot C})p^2}+\eta^2{(\mathbf{B\cdot p})}(\mathbf{C\cdot p})\right]^2\,.
\end{align}
Employing the additional restriction of parallel or antiparallel vectors, $\mathbf{C}=\alpha\mathbf{B}$
with $\alpha\in\mathbb{R}\setminus \{0\}$ providing $(\mathbf{B}\cdot\mathbf{C})^2=\mathbf{B}^2\mathbf{C}^2$ and
$\mathbf{B}(\mathbf{C}\cdot\mathbf{p})=\mathbf{C}(\mathbf{B}\cdot\mathbf{p})$, produces a nonbirefringent dispersion
relation
\begin{equation}
p_0^2=\mathbf{p}^2-(\mathbf{\hat{B}}\cdot\mathbf{p})^2-\frac{1}{2\eta^2\alpha\mathbf{B}^2}\,,\label{n6b}%
\end{equation}
with the unit vector $\mathbf{\hat{B}=B/}\vert\mathbf{B}\vert$.
Here, $\alpha>0$ stands for parallel vectors, whereas $\alpha<0$ describes antiparallel ones.
Introducing an angle via the relationship $\mathbf{B\cdot p}=\vert\mathbf{B}\vert\vert\mathbf{p}\vert\cos\vartheta$, yields:
\begin{equation}
p_{0}=\sqrt{\mathbf{p}^2\sin^2\vartheta-\frac{1}{2\eta^{2}\alpha\mathbf{B}^2}}\,.
\label{n6b2}%
\end{equation}
This dispersion relation is similar to that of Podolsky's theory, but the momentum-dependent part is
modified by the factor $\sin^2\vartheta$.
Moreover, the ``mass'' term, $1/(2\eta
^{2}\alpha\mathbf{B}^2)$, appears with a negative sign for $\alpha>0$, which jeopardizes the energy definition.
Relation~(\ref{n6b}) can be investigated with respect to causality.
\begin{figure}[ptb]
\subfloat[]{\label{caseNzero-positive-alpha}\includegraphics[scale=0.35]{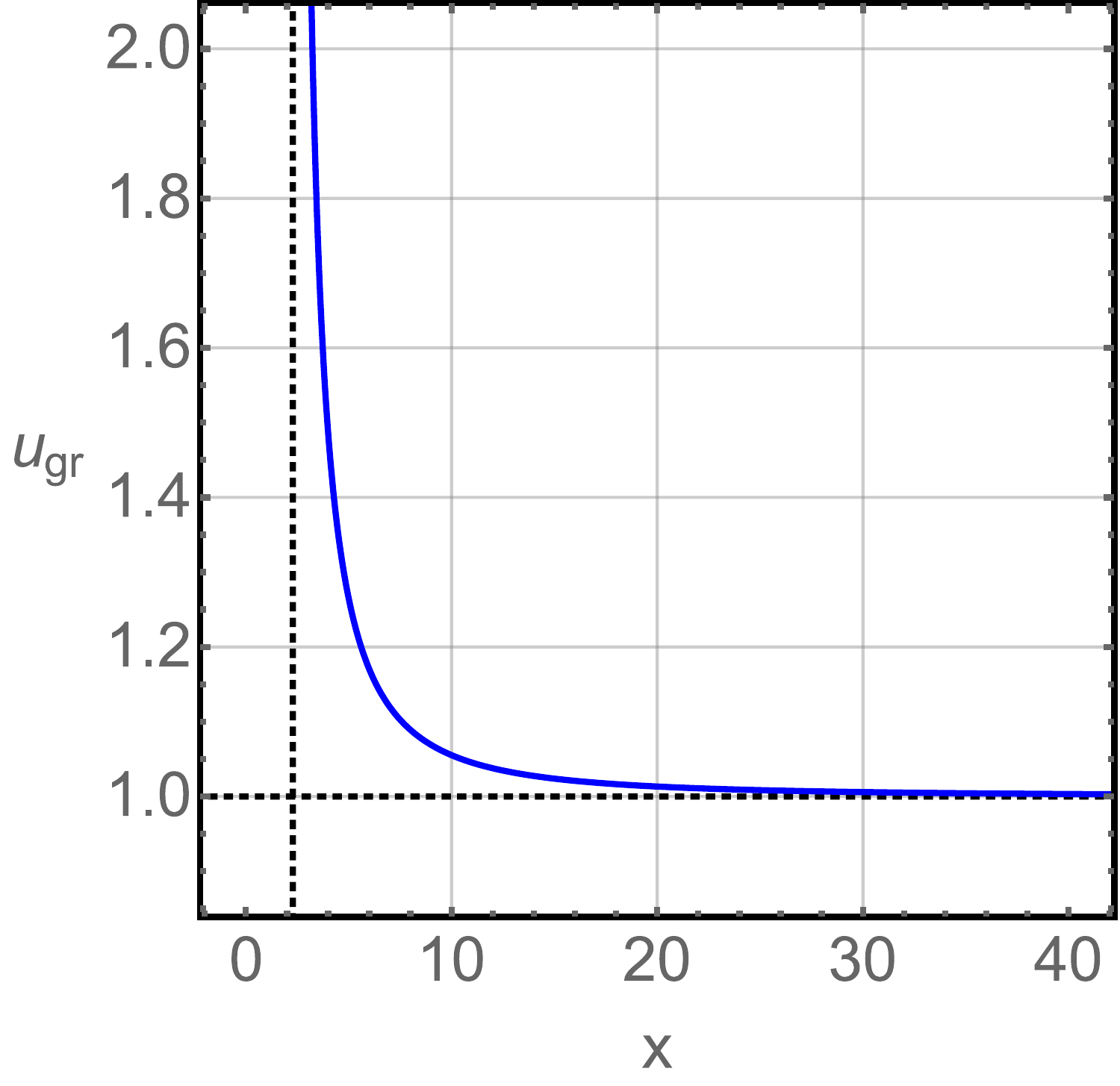}}\hspace{2cm}
\subfloat[]{\label{caseNzero-negative-alpha}\includegraphics[scale=0.35]{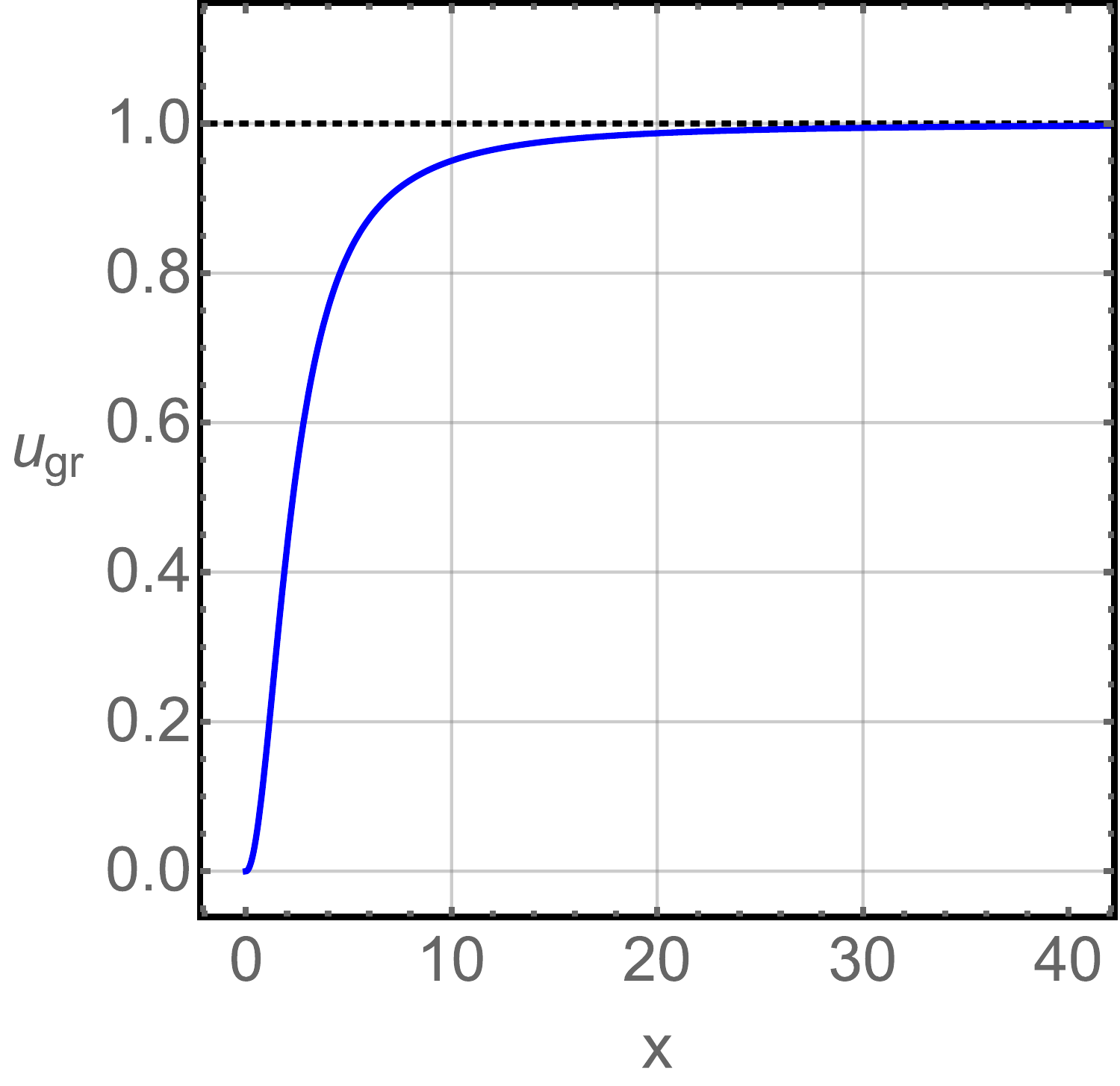}}
\caption{Group velocity of Eq.~(\ref{ug1}) for $\alpha>0$ \protect\subref{caseNzero-positive-alpha} and for
$\alpha<0$ \protect\subref{caseNzero-negative-alpha} where $\vartheta=9\pi/10$. Asymptotes are illustrated as black, dotted lines.}%
\label{caseNzero}%
\end{figure}
First, the front velocity is given by $u_{\mathrm{fr}}=|\sin\vartheta|\leq1$ independently from $\alpha$ and
does not show any vicious behavior. Second, the group velocity reads
\begin{equation}
{\mathbf{u}_{\mathrm{gr}}}=\frac{\mathbf{p}-\mathbf{\hat{\mathbf{B}}}(\mathbf{\hat{\mathbf{B}}}\cdot\mathbf{p})}{p_{0}}\,,
\end{equation}
having a modulus given by
\begin{equation}
\label{ug1}
u_{\mathrm{gr}}=\frac{1}{\sqrt{1-\mathrm{sgn}(\alpha)[2x^2\sin^2%
\vartheta]^{-1}}}\,,\quad x=\eta|\mathbf{p}|\sqrt{|\alpha|}|\mathbf{B}|\,.
\end{equation}
For $\alpha>0$, the velocity (\ref{ug1}) is only well-defined for $\mathbf{p}^{2}>{(2\eta^{2}\alpha\vert \mathbf{B}\vert ^{2}\sin
^{2}{\vartheta})^{-1}}.$ One problem is that this minimum cut-off depends on the angle $\vartheta$, which is why it does not really work as a true
cut-off. Besides, the mode does not behave like a standard photon for vanishing $\eta$. It is easy to notice that the group velocity is
always larger than 1 and even singular for $\eta\vert \mathbf{p}\vert\vert \mathbf{B}\vert=\csc{\vartheta}/\sqrt{2\alpha}$, i.e.,
the corresponding mode violates causality and is spurious. Its behavior is depicted in Fig.~\ref{caseNzero-positive-alpha}
as a function of $x$. In contrast, the group velocity for $\alpha<0$ does not
exhibit any singularities. This mode is exotic, as its group velocity increases from
zero and approaches the limit $u_{\mathrm{gr}}=1$ from below. A plot of its modulus is presented in
Fig.~\ref{caseNzero-negative-alpha}.

Another possibility worthwhile to consider is that of orthogonal $\mathbf{B}$ and $\mathbf{C}$,
$\mathbf{B\cdot C}=0$. This configuration is especially interesting, because
it implies $D_{ii}=0$ and Tr($D_{\mu\nu})=0$, whereupon it represents a
higher-derivative electrodynamics that does not contain Podolsky's sector at
all. The dispersion equation (\ref{n4}) is%
\begin{equation}
\label{n7}
p^4(\eta^4{\mathbf{B}^2\mathbf{C}^2})+\eta^4p^2\left[%
{\mathbf{B}^2(\mathbf{C\cdot p})^2}+\mathbf{C}^2(\mathbf{B\cdot p})^2\right]
-2\eta^2{(\mathbf{B\cdot p})}(\mathbf{C\cdot p})-1=0\,.
\end{equation}
To analyze this situation, we adopt a coordinate system in which the
vectors $\mathbf{C}$, $\mathbf{B}$ point along the $x$ and $y$ axis,
respectively: $\mathbf{C}={|\mathbf{C}|}\mathbf{\hat{e}}_{x}$, $\mathbf{B}%
={|\mathbf{B}|}\mathbf{\hat{e}}_{y}$. The momentum $\mathbf{p}$ shall
enclose an angle $\vartheta$ with the $z$ axis. In this system, it holds
that
\begin{equation}
\label{basis}
\mathbf{p}=\left\vert \mathbf{p}\right\vert \left(  \sin\vartheta\cos\phi
,\sin\vartheta\sin\phi,\cos\vartheta\right)  ,
\end{equation}
which implies ${\mathbf{B\cdot p}}=\left\vert \mathbf{B}\right\vert \left\vert
\mathbf{p}\right\vert \sin\vartheta\cos\phi$ and ${\mathbf{C\cdot p}%
}=\left\vert \mathbf{C}\right\vert \left\vert \mathbf{p}\right\vert
\sin\vartheta\sin\phi.$ The relation~(\ref{n7}) is rewritten as
\begin{subequations}
\begin{equation}
{(p_0^{(\pm)})^2}=\alpha_{\vartheta}\vert\mathbf{p}\vert^2\pm{\frac{1}{\eta^2{\vert\mathbf{B}\vert}\vert\mathbf{C}\vert}
\sqrt{\frac{1}{4}\nu_1^2\vert\mathbf{p}\vert^4+\nu_1\vert \mathbf{p}\vert^2\sin2\phi+1}}\,,
\end{equation}
with%
\begin{equation}
\alpha_{\vartheta}={1-\frac{1}{2}\sin^2\vartheta}\,,\quad
\nu_1=\eta^2{\vert\mathbf{B}\vert}\vert\mathbf{C}\vert\sin^2\vartheta\,.
\end{equation}
\end{subequations}
In what follows, modes denoted by the glyph $\oplus$ will be
identified with the upper sign choice of a dispersion relation, whereas modes called
$\ominus$ will correspond to the lower one.
The front velocity reads $u_{\mathrm{fr}}=1$ independently of the angles
$\vartheta$ and $\phi$, which is a result compatible with causality. The
next step is to evaluate the group velocity, whose modulus can be expressed in terms of the basis
$\mathbf{B}={|\mathbf{B}|}\mathbf{\hat{e}}_{y},\mathbf{C}={|\mathbf{C}%
|}\mathbf{\hat{e}}_{x}$ by using the representation of the momentum of
Eq.~{(\ref{basis})}:
\begin{figure}[t]
\subfloat[]{\label{spacelike-case-orthogonal-vectors-1}\includegraphics[scale=0.35]{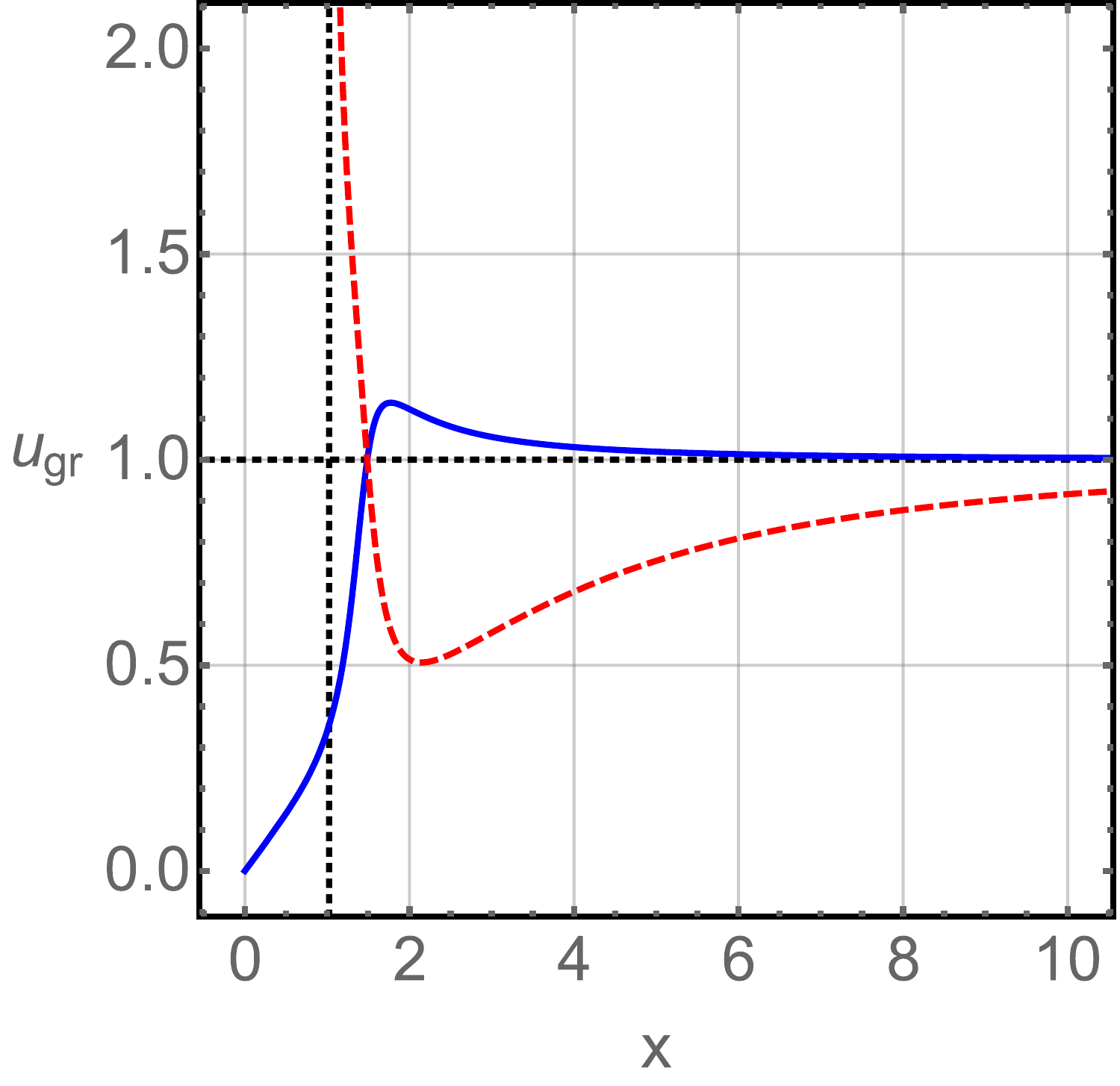}}\hspace
{2cm}
\subfloat[]{\label{spacelike-case-orthogonal-vectors-2}\includegraphics[scale=0.35]{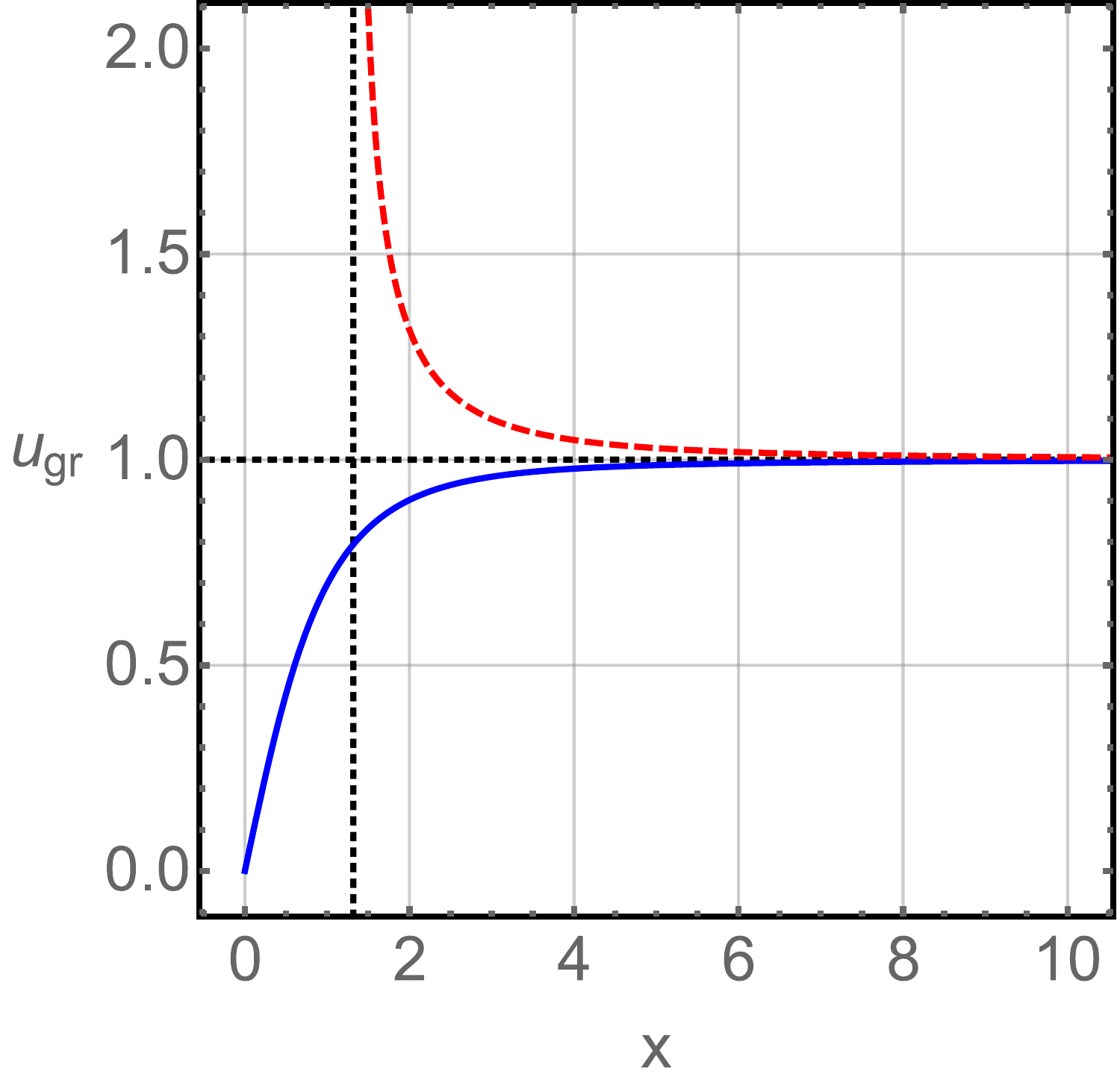}}
\caption{Group velocity {of
Eq.~(\ref{eq:group-velocity-parity-even-orthogonal})} for {$\vartheta=3\pi/7$%
}, $\phi=4\pi/5$ \protect\subref{spacelike-case-orthogonal-vectors-1} and for
{$\vartheta=3\pi/4$}, $\phi=2\pi/5$
\protect\subref{spacelike-case-orthogonal-vectors-2}. The group velocity for the mode
$\oplus$ is represented by a plain, blue line, whereas the mode $\ominus$ is
shown as a red, dashed line. Asymptotes are illustrated as black, dotted lines.}%
\label{caseN}
\end{figure}
\begin{subequations}
\label{eq:group-velocity-parity-even-orthogonal}%
\begin{equation}
\label{ug3}
{u_{\mathrm{gr}}^{(\pm)}}=\sqrt{\frac{4x^{2}\widetilde{\mathcal{S}}%
+2x^{4}\widetilde{\mathcal{T}}+x^{6}\widetilde{\mathcal{U}}}{Y^{2}%
\{x^{2}[3+\cos(2\vartheta)]\pm2Y\}}}\,,
\end{equation}
with
\begin{align}
\widetilde{\mathcal{S}} &
=3+\cos(2\vartheta)\pm Y\sin^{2}\vartheta\sin(2\phi)\,,\displaybreak[0]\\[2ex]
\widetilde{\mathcal{T}} &  =2[3+\cos(2\vartheta)]\sin^{2}\vartheta\sin
(2\phi)\pm Y\sin^{4}\vartheta\,,\displaybreak[0]\\[2ex]
\widetilde{\mathcal{U}} &  =[3+\cos(2\vartheta)]\sin^{4}\vartheta
\,,\displaybreak[0]\\[2ex]
Y &  =\sqrt{4+x^{2}\sin^{2}\vartheta\lbrack x^{2}\sin^{2}\vartheta+4\sin
(2\phi)]}\,,\quad x=\eta|\mathbf{p}|\sqrt{|\mathbf{B}||\mathbf{C}|}\,.
\end{align}
The graph of Fig.~\ref{caseN} displays the {modulus of the group velocity}
of the two modes $\oplus$, $\ominus$ considered in Eq.~(\ref{ug3}) for
different choices of the angles. Some
comments are {worthwhile}. First, we point out that the group velocity of the
mode $\ominus$ diverges for the following choice of the
dimensionless variable $x$:
\end{subequations}
\begin{equation}
x_{\mathrm{sing}}=\frac{|\sec\vartheta|}{2}\sqrt{\sqrt{[3+\cos(2\vartheta
)]^{2}-4\sin^{4}\vartheta\cos^{2}(2\phi)}+2\sin^{2}\vartheta\sin(2\phi)}\,.
\end{equation}
This value is indicated by the dashed, vertical lines in
Fig.~\ref{caseN}. Because of the singularity, the behavior of the latter mode at momenta lying in
this regime is unphysical. In contrast, the group velocity of the mode
$\oplus$ (continuous line) does not have any singularities.
For small momenta, the group velocities can be expanded as follows:
\begin{equation}
\label{eq:group-velocity-parity-even-small-momentum}
{u_{\mathrm{gr}}^{(\pm)}}=\widetilde{\Theta}^{(\pm)}|\mathbf{p}|\,,\quad \widetilde{\Theta}^{(\pm)}=\eta\sqrt{|\mathbf{B}||\mathbf{C}|}\sqrt{\frac{1}{2}[\sin(2\phi)\mp 1]\sin^2\vartheta\pm 1}\,.
\end{equation}
Each behaves like a Podolsky mode whose mass depends on the momentum direction.
Note that such expansions can also be interpreted as expansions valid for a small
controlling coefficient $\eta$.
Furthermore, Fig.~\ref{spacelike-case-orthogonal-vectors-1} shows that the
group velocity of the mode $\oplus$ has a maximum, approaches 1 from above,
and becomes larger than 1 (breaking causality) for a certain range of parameters.
For the same values of the angles, the group velocity of the mode $\ominus
$ decreases from its initial singularity to reach a minimum and finally
approaches 1 from below. The graph~\ref{spacelike-case-orthogonal-vectors-2},
generated for other values of the angles does not reveal any maxima or
minima, with the group velocity of the mode $\ominus$ approaching 1 from
above, and the mode $\oplus$ approaching 1 from below. For this parameter choice,
the mode $\oplus$ is exotic. It was also
verified that the mode $\ominus$  does not propagate for certain
angles. For example, the group velocity vanishes identically for $\vartheta=\pi/2$ and $\phi=\pi/4$
where for $\vartheta=\pi/2$ and $\phi<[\pi+\arcsin(1/x^2)]/2$ it even takes complex values.

So, we conclude that the spacelike configuration, $B_{\mu}=(0,\mathbf{B})_{\mu}$ and $C_{\mu}=(0,\mathbf{C})_{\mu}$,
with parallel or orthogonal $\mathbf{B}$ and $\mathbf{C}$ can yield both exotic and spurious dispersion relations whose group
velocities diverge or break causality. In general, such relations suggest meaningful signal
propagation only in a regime of large momenta that is compatible with {$u_{\mathrm{fr}}=1$}. However,
this interpretation is not capable of recovering a certain sector, once the dispersion relations
associated cannot be considered as physical for a given momentum range.

\subsubsection{Parity-odd anisotropic sector (with $\theta=0)$}
\label{sec:parity-odd-anisotropic}

A parity-odd anisotropic configuration is characterized by a purely timelike direction $C_{\mu}=(C_0,0)_{\mu}$ and
a purely spacelike one $B_{\mu}=(0,\mathbf{B})_{\mu}$, which leads to%
\begin{subequations}
\begin{equation}
p_{0}=\frac{\mathrm{sgn}(C_0)\left\vert \mathbf{p}\right\vert \sqrt
{\eta^4C_0^2|\mathbf{B}\times\mathbf{p}|^4+\mathbf{B}^2}
-\mathbf{B\cdot p}}{\eta^{2}C_0|\mathbf{B}\times\mathbf{p}|^2}\,,
\end{equation}
or
\begin{equation}
\label{eq:dispersion-relation-parity-odd}
p_0=\frac{\mathrm{sgn}(C_0)\sqrt{1+\eta^4C_0^2\mathbf{B}^2\mathbf{p}^4\sin^4\vartheta_B}-\cos\vartheta_B}{\eta^2C_0\vert\mathbf{B}\vert\vert\mathbf{p}\vert\sin^2\vartheta_B}\,,
\end{equation}
\end{subequations}
with ${\mathbf{B\cdot p}}=\vert\mathbf{B}\vert\vert\mathbf{p}\vert\cos\vartheta_B$,
$|\mathbf{B}\times\mathbf{p}|=|\mathbf{B}||\mathbf{p}|\sin\vartheta_B$. The dispersion relation is
positive, real-valued, and not defined in the limit $C_0\mapsto0$. The
result for the front velocity is simply $u_{\mathrm{fr}}=1,$ not revealing any
problems with causality in the large-momentum regime. The
associated group velocity, however, is%
\begin{subequations}
\begin{align}
{u_{\mathrm{gr}}} &  =\left[(\eta|\mathbf{p}|)^{-4}C_0^{-2}\Upsilon^{-3}
\left\{(\mathbf{B}\cdot\widehat{\mathbf{p}})\left[{\mathrm{sgn}%
(C_0)}2\Xi(\Upsilon+4\mathbf{B}^2)-4\mathbf{B}^2(\mathbf{B}%
\cdot\widehat{\mathbf{p}})\right]  +2\mathbf{B}^2(\Upsilon-2\mathbf{B}^2)
-4\Xi^2\Upsilon\right\}\right. \nonumber\\
&  \left.\phantom{{}={}}+5-4\Upsilon^{-1}\mathbf{B}^2\left[\mathrm{sgn}(C_{0})2\Xi^{-1}(\mathbf{B}\cdot\widehat{\mathbf{p}})-1\right]
-4(\eta|\mathbf{p}|)^4\mathbf{B}^2C_0^2\Upsilon\Xi^{-2}\right]^{1/2}\,,
\end{align}
with
\begin{equation}
\Upsilon=(\mathbf{B}\cdot\widehat{\mathbf{p}})^2-\mathbf{B}^2\,,\quad
\Xi=\sqrt{\mathbf{B}^2+(\eta|\mathbf{p}|)^{4}C_0^2\Upsilon^2}\,.
\end{equation}
\begin{figure}[ptb]
\includegraphics[scale=0.35]{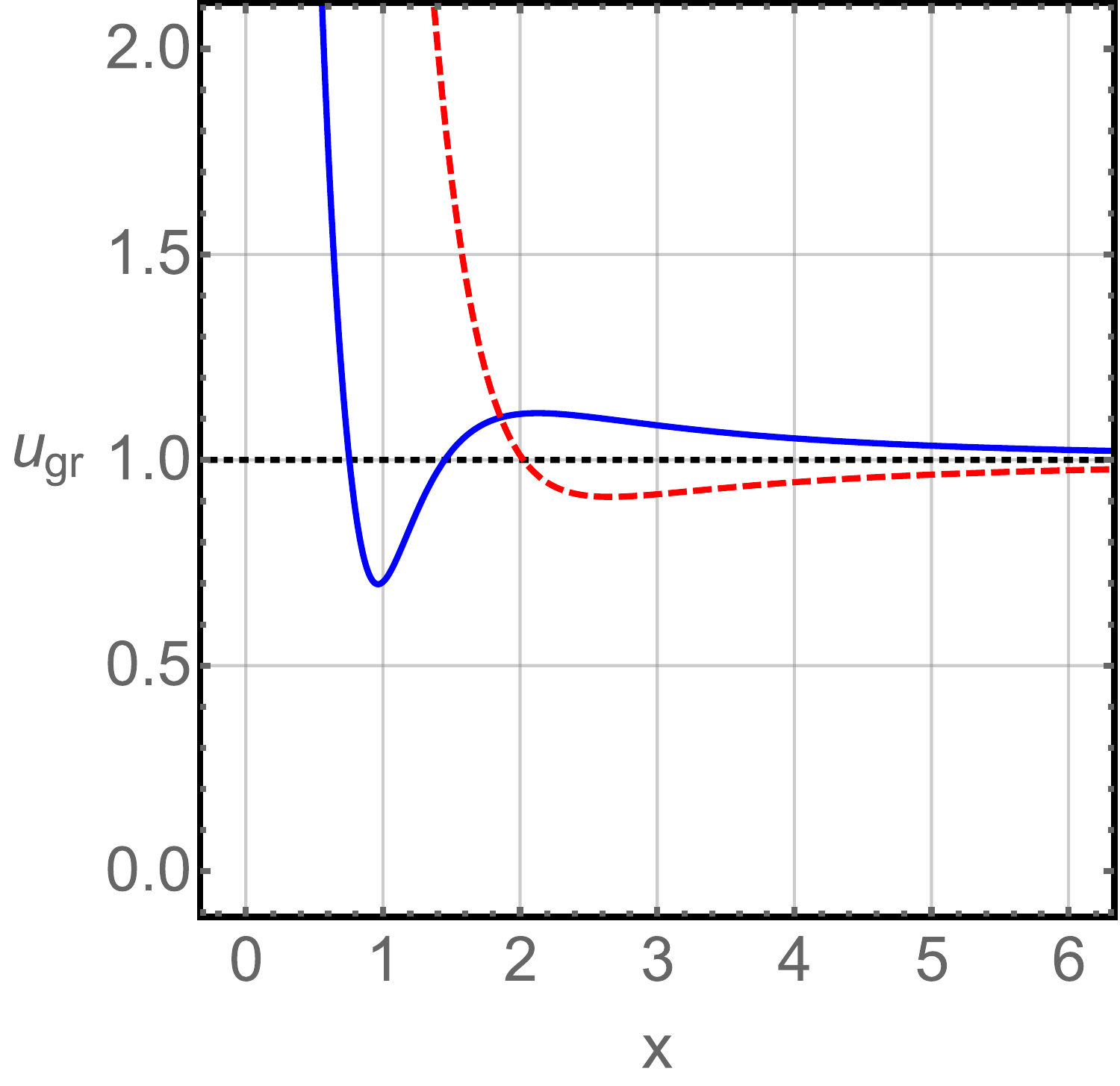}\caption{Group velocity of
Eq.~\eqref{eq:group-velocity-parity-odd-case} for $\vartheta_B=3\pi/10$,
$\mathrm{sgn}(C_0)=1$ (blue, plain) and $\mathrm{sgn}(C_0)=-1$ (red,
dashed). The horizontal asymptote is illustrated as a black, dotted line.}%
\label{caseN2}
\end{figure}
Note that there is, in principle, a single mode only, i.e., the framework
considered describes a nonbirefringent
vacuum for photons. However, both the dispersion law and the group
velocity depend on the sign of the component $C_{0}$ explicitly. In terms of
the angle $\vartheta_{B}$ between $\mathbf{B}$ and $\mathbf{p}$, the
result is expressed as
\end{subequations}
\begin{align}
\label{eq:group-velocity-parity-odd-case}
u_{\mathrm{gr}} &  =\left[1-4\csc^2\vartheta_B+{\mathrm{sgn}(C_0)}
\frac{8}{Y}\cot\vartheta_B\csc\vartheta_B+\frac{4x^4}{Y^2}\sin^2\vartheta_B\right. \nonumber\\
&\phantom{{}={}}\left.\,+\,\frac{1}{8{x^4}}\left\{\csc^6\left(\frac{\vartheta_B}{2}\right)(1-{\mathrm{sgn}(C_0)}Y)
+\sec^6\left(\frac{\vartheta_B}{2}\right)(1+{\mathrm{sgn}(C_0)}Y)\right\}\right]^{1/2}\,,
\end{align}
where $Y=\sqrt{1+{(x\sin\vartheta)^{4}}}\,,$ $x={\eta|\mathbf{p}%
|\sqrt{|\mathbf{B}||C_{0}|}}\,.$ The group velocity becomes singular for a
vanishing momentum only, see Fig.~\ref{caseN2}, which characterizes an
unphysical regime for each sign of $C_{0}$, therefore. An
expansion for a small momentum and controlling coefficient $\eta$, respectively,
provides
\begin{equation}
u_{\mathrm{gr}}=\frac{1}{2\sqrt{2}x^2}\sqrt{[1+\mathrm{sgn}(C_0)]\sec^6\left(\frac{\vartheta_B}{2}\right)+[1-\mathrm{sgn}(C_0)]\csc^6\left(\frac{\vartheta_B}{2}\right)}\,,
\end{equation}
explicitly revealing the singularities for both possible signs of $C_0$.
Note that the group velocity of one of the modes (for $C_0<0$)
has a minimum and approaches 1 from below for increasing $x$, whereas the group velocity of the
second mode (for $C_0>0$) has both a minimum and a maximum and approaches 1
from above. The modulus of the group velocity becomes
larger than 1 for both modes at some values of $x,$ which is
a behavior that corresponds to causality violation. We also point out that the
two modes merge into a single mode for
$\vartheta=\pi/2,$ and that they interchange their role for $\vartheta\in(\pi/2,\pi]$.
Both modes do not behave like standard photons when $\eta$ vanishes.
As their group velocities exhibit singularities, they are spurious.

The compilation of the results obtained seems to reveal that this
higher-derivative electrodynamics only exhibits well-behaved signal propagation in
the limit of large momenta. But, the dispersion relations cannot be valid only for a given momentum range, unless a physically reasonable cutoff can be imposed on the theory. So, this interpretation will not be considered at this stage. For small momenta, some of the modes behave like a Podolsky mode where the Podolsky parameter can depend on the direction
of the momentum due to anisotropy. We encountered exotic modes such as the dispersion relation $\oplus$ for the parameter choice used in Fig.~\ref{spacelike-case-orthogonal-vectors-2}. The others that exhibit nonphysical features must be considered
as spurious.

\subsection{Unitarity analysis}
\label{sec:unitarity-analysis}

At this point, we present some discussion on unitarity of this
higher-derivative model. The analysis of unitarity at tree-level can be
performed by means of a contraction of the propagator with external currents
\cite{Veltman}, which leads to a Lorentz scalar that is often referred to as
the saturated propagator ($\mathit{SP}$):%
\begin{equation}
\mathit{SP}\equiv J^{\mu}\Delta_{\mu\nu}J^{\nu}\,.\label{Sat2}%
\end{equation}
The gauge current $J^{\mu}$ is taken as real and satisfies the conservation law
$\partial_{\mu}J^{\mu}=0$, which in momentum space reads $p_{\mu
}J^{\mu}=0$. Therefore, a contraction with this current eliminates all
contributions whose tensor structure contains at least a single four-momentum.
The latter are associated with the gauge choice, i.e., these terms do not
describe the physics of the system under consideration. In accordance with
this method, unitarity is assured whenever the imaginary part of the residue
of the saturation $SP$ evaluated at a vanishing denominator is
positive.

We first assess the usual Podolsky theory, whose propagator is given by
Eq.~(\ref{PPodolsky}). Contractions with the external four-current provide
\begin{equation}
\mathit{SP}=-\mathrm{i}\left(  \frac{J^{\nu}\Theta_{\nu\alpha}J^{\alpha}}{p^{2}\left(
1-\theta^{2}p^{2}\right)  }-\frac{\xi}{p^{2}}J^{\nu}\Omega_{\nu\alpha}J^{\alpha}\right)\,,
\end{equation}
which simplifies to
\begin{equation}
\mathit{SP}=\frac{\mathrm{i}}{\theta^{2}}\left(  \frac{J^{2}}{p^{2}\left(  p^{2}-1/\theta
^{2}\right)  }\right)\,,
\end{equation}
due to current conservation. As properly noted in
Ref.~\cite{Turcati}, the latter can be written as%
\begin{equation}
\mathit{SP}=\mathrm{i}\left(-\frac{J^{2}}{p^{2}}+\frac{J^{2}}{p^{2}-1/\theta
^{2}}\right)\,.
\end{equation}
For a physical spacelike current, $J^{2}<0,$ the pole $p^{2}=0$ yields a
positive imaginary part of the residue of $\mathit{SP}$, while the pole $p^{2}%
=1/\theta^{2}$ provides a negative one, so that this theory reveals a
nonunitary behavior associated with the massive mode. The Lee-Wick theories
are plagued by the same problem, but mechanisms to recover unitarity by
suppressing the negative-norm states are stated in the literature
\cite{LWghost}. Within this scenario, we point out that working with a
Poldosky Lagrangian, whose term proportional to $\theta^{2}$ has a reversed sign,
does not solve the unitarity problem at all (it only moves the
problem from one pole to the other). Instead, it leads to a dispersion relation
that is not well-defined for small enough momenta.

Considering the propagator of Eq.~(\ref{PropLV6}), defined in the absence of the
Podolsky term ($\theta^{2}=0$), the saturation reads%
\begin{align}
\label{eq:saturated-propagator}
\mathit{SP} &  =-\frac{\mathrm{i}}{p^{2}\Gamma(p) }\Big\{\Gamma(
p)J^{2}-\,2\eta^{2}p^{2}(B\cdot J)(C\cdot
J)\Pi_{\theta=0}(p)\nonumber\\
&  \phantom{{}={}}\hspace{1.6cm}\left. +\,\eta^{4}p^{2}(B\cdot J) ^{2}\left[
(C\cdot p) ^{2}-C^{2}p^{2}\right] +\eta^{4}p^{2}(C\cdot J)^{2}\left[  ( B\cdot
p) ^{2}-B^{2}p^{2}\right] \right\} \,,
\end{align}
with $\Pi_{\theta=0}(p)$ given by Eq.~(\ref{P3}). The latter
should be analysed with respect to the residue of each denominator.

\begin{itemize}
\item First pole: The residue of the saturation evaluated
at $p^2=0$ is simply given by
\begin{equation}
\mathrm{Res}(\textit{SP})\vert_{p^2=0}=-\mathrm{i}J^{2}\,.
\end{equation}
A spacelike $J^{2}<0$ yields
$\mathrm{Im}[\mathrm{Res}(\textit{SP})\vert_{p^2=0}]>0$,
which is why unitarity is assured. Therefore, we deduce that a quantization of this
mode, which is the standard one of electrodynamics, corresponds to the photon.
\item Second pole: Evaluating the residue at $\Gamma (p)=0$ works
in the same manner, in principle:
\begin{align}
\mathrm{Res}(\mathit{SP})|_{\Gamma (p)=0}& =-\mathrm{i}\eta^2\Big\{\eta
^{2}(B\cdot J)^{2}\left[ (C\cdot p)^{2}-C^{2}p^{2}\right] +\eta ^{2}(C\cdot
J)^{2}\left[ (B\cdot p)^{2}-B^{2}p^{2}\right]   \notag \\
& \phantom{{}={}}\hspace{1.0cm}-2(B\cdot J)(C\cdot J)\Pi _{\theta =0}(p)%
\Big\}\Big|_{\Gamma (p)=0}\,.
\end{align}%
The situation now is more involved, and requires a careful analysis to find
the global sign of the expression in curly brackets. Considering Eq.~(\ref%
{lambp}), an evaluation of $\Gamma (p)=0$ provides the useful relationship%
\begin{equation}
\mathrm{sgn}(\Pi_{\theta=0})\Pi (p)|_{\substack{ \theta =0 \\ \Gamma (p)=0}}=\eta ^{2}%
\sqrt{\left[ (B\cdot p)^{2}-B^{2}p^{2}\right] \left[ (C\cdot
p)^{2}-C^{2}p^{2}\right] }\big|_{\Gamma (p)=0}\,.
\label{eq:definitions-gamma-pi}
\end{equation}%
The sign function on the left-hand side must be taken into account, as $\Pi
_{\theta =0}(p)$ can be a negative quantity onshell, whereas the right-hand
side is manifestly positive. Furthermore, we again have to distinguish
between two cases. After investigating a large number of Lorentz-violating
sectors numerically, we deduced that either $p^{\mu }\in M_{1}$ or $p^{\mu
}\in M_{2}$ with the two sets $M_{1}\equiv \{(B\cdot
p)^{2}-B^{2}p^{2}>0,(C\cdot p)^{2}-C^{2}p^{2}>0\}$ or $M_{2}\equiv \{(B\cdot
p)^{2}-B^{2}p^{2}<0,(C\cdot p)^{2}-C^{2}p^{2}<0\}$. Both options
assure a real $\Pi(p).$ For the first option, the residue can be
expressed in terms of the square of a sum of two terms. For the
second option, a global minus sign must be pulled out of the whole
expression, reversing the global saturation sign. The residues can then
neatly be written as follows:
\begin{equation}
\mathrm{Res}(\mathit{SP})|_{\Gamma (p)=0}=\mathrm{i}\eta^4\left\{
\begin{array}{lcl}
-\Big\{(B\cdot J)\sqrt{(C\cdot p)^{2}-C^{2}p^{2}} & \text{for} & p^{\mu }\in
M_{1}\,, \\
\phantom{{}}-\mathrm{sgn}(\Pi_{\theta=0})(C\cdot J)\sqrt{(B\cdot p)^{2}-B^{2}p^{2}}%
\Big\}^{2}\Big|_{\Gamma (p)=0} &  &  \\
\Big\{(B\cdot J)\sqrt{|(C\cdot p)^{2}-C^{2}p^{2}|} & \text{for} & p^{\mu
}\in M_{2}\,. \\
\phantom{{}}+\mathrm{sgn}(\Pi_{\theta=0})(C\cdot J)\sqrt{|(B\cdot p)^{2}-B^{2}p^{2}|}%
\Big\}^{2}\Big|_{\Gamma (p)=0} &  &  \\
\end{array}%
\right.
\end{equation}%
\end{itemize}
Hence, $\mathrm{Im}[\mathrm{Res}(\mathit{SP})|_{\Gamma (p)=0}]<0$
for the first case, which is a behavior indicating a breakdown of unitarity.
The second case behaves in the opposite way, whereupon unitarity is preserved.
The parity-odd configuration with $(C_{\mu})=(C_0,\mathbf{0})$, $(B_{\mu})=(0,\mathbf{B})$
analysed in Sec.~\ref{sec:parity-even-anisotropic} is covered by the latter case. We find
\begin{equation}
(B\cdot p)^{2}-B^{2}p^{2}=-\frac{|\mathbf{B}|}{2\eta ^{2}|\mathbf{C}|}%
<0\,,\quad (C\cdot p)^{2}-C^{2}p^{2}=-\frac{|\mathbf{C}|}{2\eta ^{2}|\mathbf{%
B}|}<0\,,
\end{equation}%
which is why $p^{\mu}\in M_{2}$ in this sector. Thus, unitarity is preserved for this
configuration.

For the parity-odd case introduced in Sec.~\ref%
{sec:parity-odd-anisotropic} it holds that $(C\cdot
p)^{2}-C^{2}p^{2}=C_{0}^{2}\mathbf{p}^{2}>0$. However, it is a bit more
involved to evaluate the second condition. The inequality $(B\cdot
p)^{2}-B^{2}p^{2}\geq 0$ to be checked can be cast into a more transparent
form as follows:
\begin{subequations}
\begin{equation}
2x^4\sin ^{4}\vartheta _{B}\cos ^{2}\vartheta _{B}+\left\{ 3+\cos
(2\vartheta _{B})-4\,\mathrm{sgn}(C_{0})\cos (\vartheta _{B})\sqrt{%
1+x^4\sin ^{4}\vartheta _{B}}\right\} \geq 0\,,
\end{equation}%
which is equivalent to
\begin{equation}
2\left\{1-\mathrm{sgn}(C_{0})\cos \vartheta _{B}\sqrt{1+x^4\sin
^{4}\vartheta _{B}}\right\} ^{2}\geq 0\,.
\end{equation}%
\end{subequations}
The latter involves the angle $\vartheta _{B}$ defined directly below Eq.~(%
\ref{eq:dispersion-relation-parity-odd}) and the dimensionless quantity $x$
defined under Eq.~(\ref{eq:group-velocity-parity-odd-case}). For $C_{0}>0$, the
equality sign is valid for $\vartheta _{B}=0$ only, whereas for $C_{0}<0$ it
is valid for $\vartheta _{B}=\pi $. Apart from these very special values at
the boundaries of the interval for $\vartheta _{B}$, the condition $(B\cdot
p)^{2}-B^{2}p^{2}>0$ is fulfilled manifestly, whereby $p^{\mu }\in M_{1}$.
Therefore, according to our criterion, there are issues with unitarity for
this sector.

There is an alternative possibility of calculating the saturated
propagator. Using the transverse~(\textit{T}) and longitudinal~(\textit{L}) projection operators
$\Theta_{\mu\nu}$ and $\Omega_{\mu\nu}$ of Eq.~(\ref{eq:projectors}) transformed
to momentum space, the propagator can be decomposed into the four parts
$\Delta^{TT}_{\mu\nu}\equiv \Theta_{\mu\alpha}\Delta^{\alpha\beta}\Theta_{\beta\nu}$,
$\Delta^{TL}_{\mu\nu}\equiv \Theta_{\mu\alpha}\Delta^{\alpha\beta}\Omega_{\beta\nu}$,
$\Delta^{LT}_{\mu\nu}\equiv \Omega_{\mu\alpha}\Delta^{\alpha\beta}\Theta_{\beta\nu}$,
and $\Delta^{LL}_{\mu\nu}\equiv \Omega_{\mu\alpha}\Delta^{\alpha\beta}\Omega_{\beta\nu}$.
After performing contractions with the conserved current, only $\Delta^{TT}_{\mu\nu}$
will survive. Following this procedure, the saturated propagator can be cast into
the form
\begin{subequations}
\begin{align}
\textit{SP}&=J^{\mu}\Delta^{TT}_{\mu\nu}J^{\nu}\,,\quad \Delta^{TT}_{\mu\nu}=-\mathrm{i}\frac{(M^{-1})_{\mu\nu}}{p^2}\,, \\[2ex]
M^{\mu\nu}&=\left[(1-\theta^2p^2)g-2\eta^2D^{TT}p^2\right]^{\mu\nu}\,,\quad (D^{TT})_{\mu\nu}\equiv \Theta_{\mu\alpha}D^{\alpha\beta}\Theta_{\beta\nu}\,.
\end{align}
\end{subequations}
Hence, the behavior of unitarity is completely controlled by the totally transverse
part of the matrix $D_{\mu\nu}$.
Performing a formal partial-fraction decomposition, the standard pole $p^2=0$ can be
separated from the remaining expressions involving the Podolsky parameter and the
Lorentz-violating coefficient:
\begin{equation}
\label{eq:propagator-transverse-transverse}
\Delta_{\mu\nu}^{TT}=-\mathrm{i}\left\{\frac{g_{\mu\nu}}{p^2}+(\theta^2\delta+2\eta^2D^{TT})_{\mu}^{\phantom{\mu}\alpha}(M^{-1})_{\alpha\nu}\right\}\,.
\end{equation}
Using the explicit decomposition of $D_{\mu\nu}$ in terms of two four-vectors proposed
in Eq.~(\ref{Presc1}) and setting $\theta=0$, we can deduce that $\det(M)=\Gamma(p)$
with $\Gamma(p)$ from Eq.~(\ref{eq:definitions-gamma-pi}) for $\theta=0$. The result of Eq.~(\ref{eq:propagator-transverse-transverse})
is suitable to investigate unitarity for the case of a nonzero $\theta$ and a general
$D_{\mu\nu}$ whose form does not rely on a decomposition into two four-vectors.
The analysis of such cases is an interesting open problem.

To conclude this section, unitarity violation seems to be connected to
the appearance of additional time derivatives in the Lorentz-violating contributions of the Lagrange
density, cf.~Eqs.~(\ref{eq:lagrange-density-bilinear}), (\ref{OLV1}). A behavior of this kind is expected.
For a Lorentz-invariant, higher-derivative theory it was observed in \cite{LeeWick}, amongst other works.
The authors of~\cite{Colladay:2001wk} pointed out that additional time derivatives in the minimal SME fermion
sector lead to issues with the time evolution of asymptotic states. Related problems in the nonminimal SME
were found in the third and second paper of \cite{Kostelec1} and~\cite{Kostelec2}, respectively.

\section{Generalized model involving anisotropic Podolsky and Lee-Wick
terms}

There are two other \textit{CPT}-even dimension-6 terms
endowed with two additional derivatives, besides the LV modification~(\ref{TADLV1}),
that can be expressed in terms of the electromagnetic field strength tensor and the
tensor $D^{\mu\nu},$ namely:%
\begin{equation}
{D^{\alpha\beta}}F_{\mu\nu}\partial_{\alpha}\partial_{\beta}F^{\mu}{}^{\nu}\,,\quad
{D^{\mu\nu}}\partial_{\sigma}F^{\sigma\lambda}{}\partial_{\mu}F_{\nu\lambda}\,.
\label{LVSLAD2B}%
\end{equation}
The first one is a kind of anisotropic Lee-Wick term, while the second yields
a bilinear contribution similar to it, but with the opposite sign, so that
only one of these terms will be considered. The very same correspondence
observed for the case of the anisotropic Podolsky term holds for the anisotropic
Lee-Wick term, too: for the configuration of
a diagonal tensor $D_{\beta\alpha}$ of the form $D_{\beta\alpha}=D_{00}g_{\beta\alpha}$, it becomes
proportional to the usual Lee-Wick term, that is,
\begin{equation}
{D^{\alpha\beta}}F_{\mu\nu}\partial_{\alpha}\partial_{\beta}F^{\mu\nu}=D_{00}(F_{\mu\nu}\square F^{\mu\nu})\,.
\end{equation}
The behavior of the coefficients of the tensor $D_{\beta\alpha}$ under
discrete \textit{C}, \textit{P}, and \textit{T} operations, described in Tab.~\ref{tab:transformation-properties}, does not depend on the way
how the tensor is coupled to the electromagnetic field, being equally
valid for the anisotropic Lee-Wick structures of Eq.~(\ref{LVSLAD2B}).

In principle, the most general LV dimension-6
electrodynamics, modified by a rank-2 symmetric tensor, $D_{\beta\alpha},$
also includes the second of the anisotropic Lee-Wick contributions of Eq.~(\ref{LVSLAD2B}),
and is represented by the following Lagrangian:%
\begin{equation}%
\mathcal{L}%
=-\frac{1}{4}F_{\mu\nu}F^{\mu}{}^{\nu}+\frac{\theta^{2}}{2}\partial_{\alpha}F^{\alpha}{}^{\beta}\partial_{\lambda}F_{\phantom{\lambda}\beta}^{\lambda}
+\eta_{1}^{2}D_{\beta\alpha}\partial_{\sigma}F^{\sigma}{}^{\beta}\partial_{\lambda}F^{\lambda\alpha}
+\eta_2^2D^{\beta\alpha}\partial_{\sigma}F^{\sigma}{}^{\lambda}\partial_{\beta}F_{\alpha\lambda}
+\frac{1}{2\xi}(\partial_{\mu}A^{\mu})^2\,,\label{L3}%
\end{equation}
with $\theta\geq 0$, $\eta_1\geq 0$, and $\eta_2\geq 0$. The latter
incorporates the standard Podolsky term, its LV modification considered earlier, and the LV anisotropic Lee-Wick contribution.
Such a Lagrangian can be written in the form $\mathcal{L}=(1/2)A^{\mu}\bar{O}_{\mu\nu}A^{\nu}$ where%
\begin{align}
\bar{O}_{\mu\nu} &  =(1+\theta^2\square+2\eta_2^2{D^{\beta\alpha}}\partial_{\beta}\partial_{\alpha})\square
\Theta_{\mu\nu}+\left(2\eta_1^2D^{\beta\alpha}\partial_{\beta}\partial_{\alpha}-\frac{1}{\xi}\right)\square\Omega_{\mu\nu}+2\eta_1^2
{D_{\nu\mu}}\square^2\nonumber\\
&\phantom{{}={}}-2\eta_1^2{D_{\nu\alpha}}\square\partial_{\mu}\partial^{\alpha}
-2\eta_1^2{D_{\sigma\mu}}\square\partial_{\nu}\partial^{\sigma}\,.
\end{align}
Based on the prescription (\ref{Presc1}) for the symmetric tensor $D_{\beta\alpha}$ we obtain
\begin{align}
\bar{O}_{\mu\nu}&=(1+\theta^2\square+2\eta_2^2{\kappa\rho})\square\Theta_{\mu\nu}
+\left(2\eta_1^2{\kappa\rho}-\frac{1}{\xi}\right)\square\Omega_{\mu\nu}
+\eta_1^2{(B_{\mu}C_{\nu}+B_{\nu}C_{\mu})}\square^2\nonumber\\
&\phantom{{}={}}-\eta_1^2{(C_{\nu}\partial_{\mu}+C_{\mu}\partial_{\nu})\kappa}\square
-\eta_1^2{(\partial_{\nu}B_{\mu}+\partial_{\mu}B_{\nu})\rho}\square  \,,\label{prop1}%
\end{align}
with $\kappa$ and $\rho$ of Eq.~(\ref{karo}). Using the same tensor
algebra as that of the first case, the following propagator can be derived:
{\begin{subequations}
\begin{align}
\Delta_{\nu\alpha}& =-\frac{\mathrm{i}}{p^2\Delta(p)}\left\{\tilde{\Gamma}(p)\Theta_{\nu\alpha}
+\left[b^{\prime}-\xi\Delta(p)\right]\Omega_{\nu\alpha}\right. \nonumber\\
&\phantom{{}={}-\mathrm{i}p^2\Delta(p)}\left.-\,\mathrm{i}\tilde{F}(p)(B_{\nu}p_{\alpha}+B_{\alpha}p_{\nu})
-2\eta_1^2D_{\nu\alpha}p^2\tilde{\Pi}(p)-\mathrm{i}\tilde{H}(p)(C_{\nu}p_{\alpha}+C_{\alpha}p_{\nu})\right. \nonumber\\[1ex]
&\phantom{{}={}-\mathrm{i}p^2\Delta(p)}
\left.+\,\eta_1^4B_{\nu}B_{\alpha}\left[(C\cdot p)^2-C^2p^2\right]p^2+\eta_1^4C_{\nu}C_{\alpha}\left[(B\cdot p)^2-B^2p^2\right]p^2\right\}\,,
\end{align}
where
\begin{align}
\Delta(p) &= [1-\theta^2p^2-2\eta_2^2(B\cdot p)(C\cdot p)]\tilde{\Gamma}(p)\,, \displaybreak[0]\\[2ex]
\tilde{\Gamma}(p)& =\eta_1^4\left[(B\cdot p)^2-B^2p^2\right]\left[(C\cdot p)^2-C^2p^2\right]-\tilde{\Pi}^2(p)\,,\displaybreak[0]\\[2ex]
\tilde{\Pi}(p) &  =1-\theta^2p^2-\eta_1^2p^2(B\cdot C)+[\eta_1^2-2\eta_2^2](B\cdot p)(C\cdot p)\,, \displaybreak[0]\\[2ex]
\tilde{F}(p)& =  F(p)|_{\eta=\eta_1}\,,\quad \tilde{H}(p)=H(p)|_{\eta=\eta_1}\,,
\end{align}
\end{subequations}
with $F(p)$ and $H(p)$ of Eqs.~(\ref{Fp}) and (\ref{Hp}), respectively. The new dispersion equations read
\begin{subequations}
\begin{align}
0&=p^2[1-\theta^2p^2-2\eta_2^2(B\cdot p)(C\cdot p)]\,,\label{PDR6A}\\[2ex]
0&=\tilde{\Gamma}(p)\,.\label{PDR6B}%
\end{align}
\end{subequations}}
We now observe that in contrast to Eq.~(\ref{n3a}), Podolsky's dispersion relation is also modified, as shown in
Eq.~(\ref{PDR6A}).

\subsection{{Dispersion relations}}

To analyse the dispersion equations~(\ref{PDR6A}) and (\ref{PDR6B}), we
consider the main two background configurations -- timelike and spacelike --
discussed in the first model. The Podolsky parameter $\theta$ will not be discarded.

\subsubsection{{Timelike isotropic sector (with $\theta\neq0)$}}

The timelike isotropic configuration is characterized by $B_{\mu}=(B_0,\mathbf{0})_{\mu}$,
$C_{\mu}=(C_0,\mathbf{0})_{\mu}$. As already mentioned, the dispersion equation~(\ref{PDR6A})
shows that the usual Podolsky dispersion
relation is now modified by a contribution resulting from the
anisotropic Lee-Wick term. In this case, the dispersion relation obtained from Eq.~(\ref{PDR6A}) takes the
simple form,
\begin{equation}
p_0=\sqrt{\frac{1}{\varepsilon}\left(\mathbf{p}^2+\frac{1}{\theta^2}\right)}\,,
\label{DR51}%
\end{equation}
with $\varepsilon=1+2\eta_2^2B_0C_0/\theta^2$.
The latter is a Podolsky-like dispersion relation, modified by a kind of dielectric constant,
$\varepsilon$. For $B_0C_0>0$, the front velocity is less than one,
$u_{\mathrm{fr}}<1,$ and the group velocity,
\begin{equation}
u_{\mathrm{gr}}=\frac{\vert\mathbf{p}\vert}{\sqrt{\varepsilon}\sqrt{\mathbf{p}^2+1/\theta^2}}\,,
\end{equation}
is always less than 1, as well, ensuring the validity of causality for this
configuration. The DR~(\ref{DR51}) only makes sense in the presence of the
Podolsky term ($\theta^2\neq0)$. Note that $\theta^2$ appears
inside $\varepsilon$, as well. Now, we analyse the dispersion equation~(\ref{PDR6B})
for {the} isotropic configuration with $B_{\mu}=(B_0,\mathbf{0})_{\mu}$,
$C_{\mu}=(C_0,\mathbf{0})_{\mu}$, which can be written as
\begin{subequations}
\begin{equation}
E^{(\pm)}=\psi^{(\pm)}|\mathbf{p}|\,,\quad\psi^{(\pm)}=\psi^{(\pm)}%
(|\mathbf{p}|)\,,\label{eq:group-velocity-purely-timelike-case}%
\end{equation}
where
\begin{align}
\psi^{(+)} &  =\sqrt{\frac{1+\theta^2|\mathbf{p}|^2}{2y^2+\theta^2|\mathbf{p}|^2}}\,,\\[2ex]
\psi^{(-)} &  =\sqrt{\frac{1+2x^2+\theta^2|\mathbf{p}|^2}{2y^2+\theta^2|\mathbf{p}|^2}}\,,
\end{align}
and $x=\sqrt{B_0C_0}\eta_1|\mathbf{p}|$, $y=\sqrt{B_{0}C_{0}}\eta_{2}|\mathbf{p}|$ are
dimensionless coefficients. Note that both $x$ and
$y$ are linear functions of the three-momentum magnitude where the ratios
$x/(\theta|\mathbf{p}|)=\sqrt{B_0C_0}\eta_1/\theta\equiv\alpha$ and
$y/(\theta|\mathbf{p}|)=\sqrt{|\mathbf{B}||\mathbf{C}|}\eta_2/\theta
\equiv\beta$ are dimensionless. Hence, it is reasonable to investigate the
expressions for the characteristic velocities after replacing $x$ by $\alpha\theta|\mathbf{p}|$
and $y$ by $\beta\theta|\mathbf{p}|$, respectively. Now, the front velocities for both
modes are simply given by
\end{subequations}
\begin{subequations}
\begin{align}
u_{\mathrm{fr}}^{(+)} &  =\frac{1}{\sqrt{1+2\beta^2}}\,,\displaybreak[0]\\[2ex]
u_{\mathrm{fr}}^{(-)} &  =\sqrt{\frac{1+2\alpha^2}{1+2\beta^2}}\,.
\end{align}
\begin{figure}[ptb]
\centering\subfloat[]{\label{fig:full-theory-complicated-dispersion-equation-timelike-case}\includegraphics[scale=0.35]{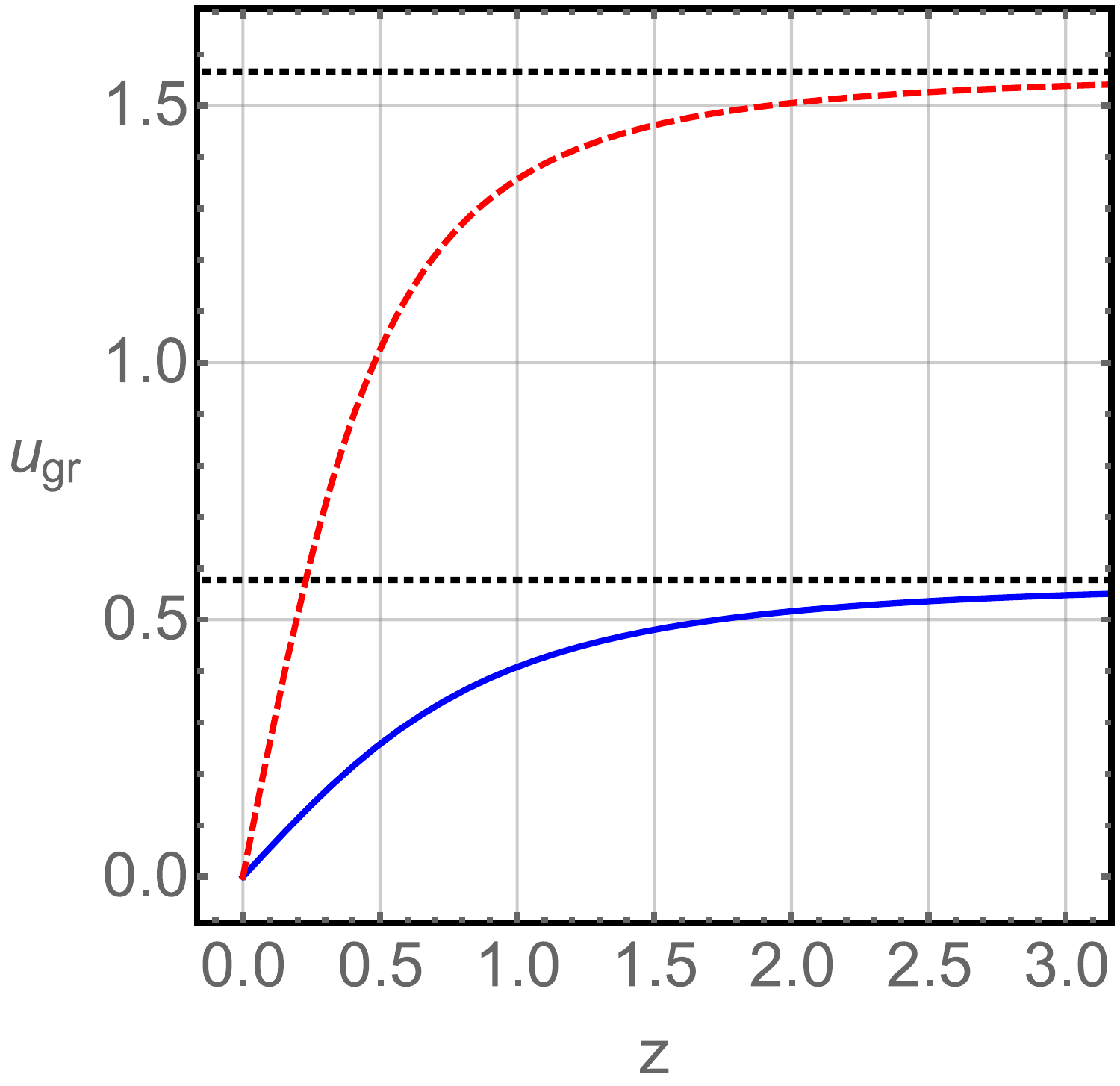}}
\caption{Group velocities of
Eq.~\eqref{eq:group-velocities-timelike-isotropic-theta} as a function of $z=\theta|\mathbf{p}|$ for the mode $\oplus$ (blue, plain) with $\beta=1$ and
for $\ominus$ with $\alpha=1$, $\beta=1/3$ (red, dashed). The horizontal asymptotes are shown as black, dotted lines.}%
\label{caseN3}%
\end{figure}
The first of these expressions is always smaller than 1, whereas the second
can be larger than 1 for $\alpha>\beta$. As the case under consideration is
isotropic, it is not too involved to obtain the group velocities:
\end{subequations}
\begin{subequations}
\label{eq:group-velocities-timelike-isotropic-theta}%
\begin{align}
u_{\mathrm{gr}}^{(+)} &  =\frac{\psi^{(+)}}{1+(\theta|\mathbf{p}|)^{-2}}\,,\\[2ex]
u_{\mathrm{gr}}^{(-)} &  =\psi^{(-)}\left\{1-\frac{1}{(\psi^{(-)})^2\left[2y^2+(\theta|\mathbf{p}|)^2\right]}\right\}\,.
\end{align}
\end{subequations}
Both expressions do not exhibit any singularities.
The graph of Fig.~\ref{caseN3} reveals decreasing group velocities
for declining momenta where each mode behaves as a Podolsky mode in this regime:
\begin{subequations}
\begin{align}
u_{\mathrm{gr}}^{(\pm)}&=\widehat{\Theta}^{(\pm)}|\mathbf{p}|\,, \\[2ex]
\widehat{\Theta}^{(+)}&=\frac{\theta}{\sqrt{1+2\beta^2}}\,,\quad\widehat{\Theta}^{(-)}=\frac{(1+2\alpha^2)\theta}{\sqrt{1+2\beta^2}}\,.
\end{align}
\end{subequations}
As the case under consideration is isotropic, the Podolsky parameters do not depend
on the momentum direction, but they involve the Lorentz-violating coefficients. Causality
violation can occur for the mode $\ominus$ dependent on the chosen parameters. The group
velocity for each mode rises from 0 to a finite value that is given by the previously
obtained expressions for the front velocity:
\begin{equation}
\lim_{z\mapsto\infty}u_{\mathrm{gr}}^{(+)}=u_{\mathrm{fr}}^{(+)}\,,\quad
\lim_{z\mapsto\infty}u_{\mathrm{gr}}^{(-)}=u_{\mathrm{fr}}^{(-)}\,.
\end{equation}
Hence, the group velocity of the mode $\ominus$ can reach values larger than
1, breaking causality, which characterizes a spurious mode. On the other hand,
the mode $\oplus$ mode does not violate causality, although it does not behave like a standard
photon for vanishing $\eta_1$ or $\eta_2$. So it is an exotic mode.
Note that vanishing Lorentz violation translates
into the limits $\alpha\mapsto 0$ and $\beta\mapsto 0$, which provides the
Podolsky-type dispersion law.

\subsubsection{{Parity-even anisotropic sector (with $\theta\neq0)$}}

For the spatially anisotropic configuration, $B_{\mu}=(0,\mathbf{B})_{\mu}$, $C_{\mu}=(0,\mathbf{C})_{\mu}$,
we can express the dispersion relation following from Eq.~(\ref{PDR6A}) as follows:%
\begin{equation}
p_0^2=\frac{1}{\theta^2}+\mathbf{p}^2-{\frac{2\eta_2^2}{\theta^2}}
(\mathbf{B\cdot p})(\mathbf{C\cdot p})\,,
\label{DR52A}%
\end{equation}
which for $\eta_2\mapsto 0$ recovers Podolsky's dispersion
relation. To analyze the dispersion relation for $\eta _{2}\neq 0$, we
first consider the situation in which the vectors $\mathbf{C}$, $\mathbf{B}$
are orthogonal. Using the coordinate system employed in Sec.~\ref{sec:parity-even-anisotropic} with the
momentum of Eq.~(\ref{basis}), implies
\begin{equation}
p_0=\sqrt{\frac{1}{\theta^2}+\mathbf{p}^2\left(1-\frac{\eta_2^2}{\theta^2}
\vert\mathbf{B}\vert\vert\mathbf{C}\vert\sin^2\vartheta\sin2\phi\right)}\,.
\label{DR52B}%
\end{equation}
\begin{figure}[b]
\centering
\includegraphics[scale=0.35]{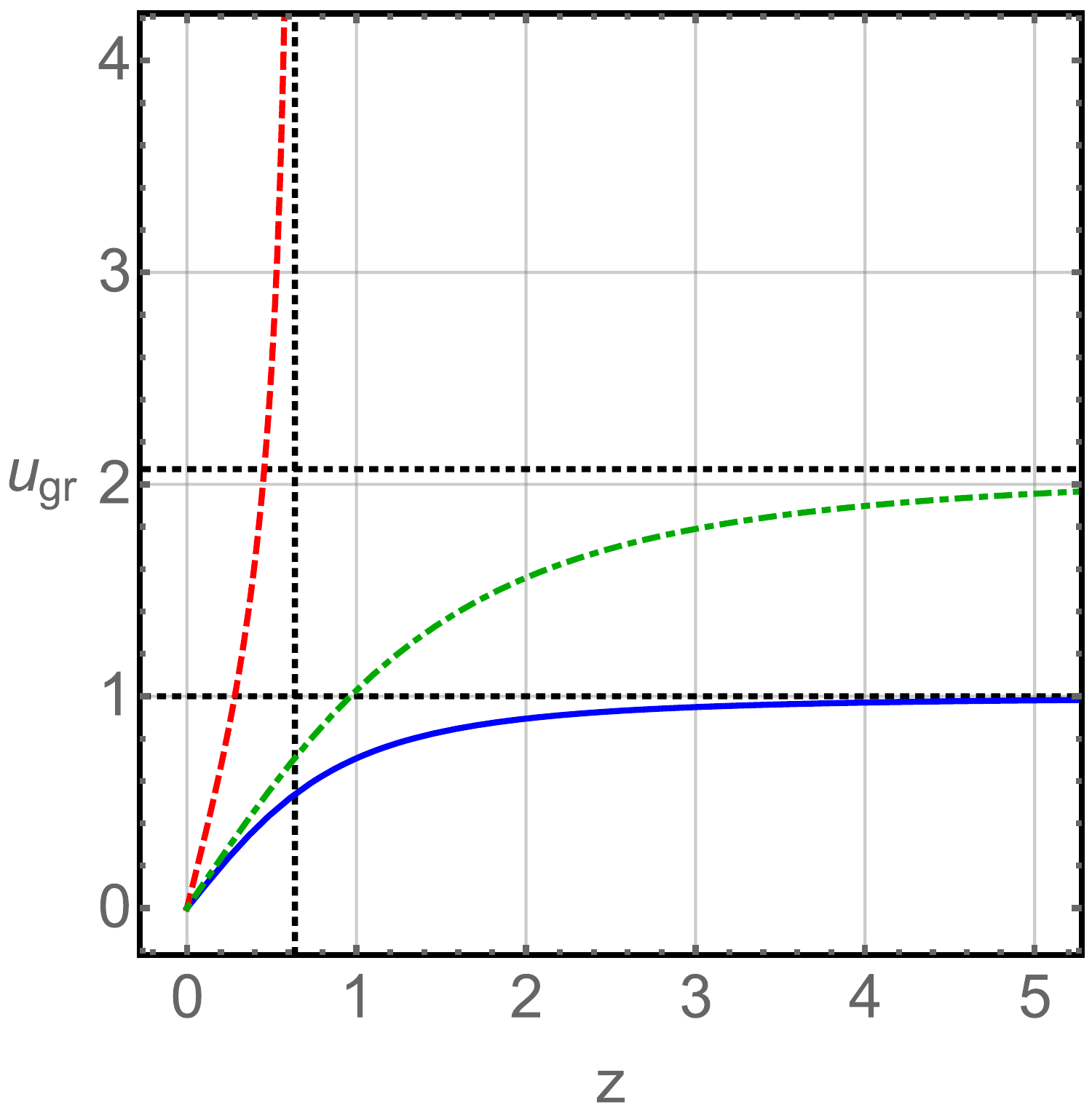}\caption{Group
velocity of Eq.~\eqref{eq:group-velocity-full-parity-pdd} for $\beta=1$,
$\vartheta=\phi=0$ (blue, plain), for $\beta=3/2$, $\vartheta=\pi/5$, $\phi=\pi/3$
(green, dash-dotted), and for $\beta=2$, $\vartheta=\pi/2$, $\phi=\pi/3$ (red,
dashed). The asymptotes are indicated by black, dotted lines.}%
\label{caseparityodd}%
\end{figure}
Here, the front velocity is%
\begin{equation}
u_{\mathrm{fr}}=\sqrt{1-{\frac{\eta_2^2}{\theta^2}\vert\mathbf{B}\vert\vert\mathbf{C}\vert}
\sin^2\vartheta\sin(2\phi)}\,,
\end{equation}
which can be larger than 1 for some values of $\phi.$
The modulus of the group velocity is given by
\begin{equation}
{u_{\mathrm{gr}}=\sqrt{2+\frac{y^4\sin^2\vartheta-z^2(2+z^2)}%
{z^2\left[1+z^2-y^2\sin^2\vartheta\sin(2\phi)\right]}}}\,,
\label{eq:group-velocity-full-parity-pdd}%
\end{equation}
with $y=\sqrt{|\mathbf{B}||\mathbf{C}|}\eta_{2}|\mathbf{p}|$,
$z=\theta|\mathbf{p}|$. Here we introduce the constant ratio $y/z=\sqrt
{|\mathbf{B}||\mathbf{C}|}\eta_{2}/\theta\equiv\beta$. The graph of
Fig.~\ref{caseparityodd} shows the behavior of the group velocity for
distinct values of the parameter $\beta$ and the angles. The latter $\beta$
can be chosen such that $u_{\mathrm{gr}}$ rises steadily from 0 and approaches
a certain constant from below for increasing $z$. Explicitly, this constant
exceeds 1 at
\begin{equation}
z|_{u_{\mathrm{gr}}=1}=\frac{\csc\vartheta}{|\beta|\sqrt{\beta^{2}%
-\sin(2\phi)}}\,,
\end{equation}
whereupon there will be issues with causality.
This value of $z$ is real for $\beta>\sqrt{\sin(2\phi)}$ only, which
demonstrates that causality problems do not arise necessarily.
Besides, $u_{\mathrm{gr}}$ also exhibits a singularity for a suitable
choice of the parameters that lies at the value
\begin{equation}
z_{\mathrm{sing}}=\frac{1}{\beta^{2}\sin^{2}\vartheta\sin(2\phi)-1}%
\,,
\end{equation}
revealing an unphysical regime.
For small momenta, the group velocity of Eq.~(\ref%
{eq:group-velocity-full-parity-pdd}) behaves as
\begin{equation}
u_{\mathrm{gr}}=\breve{\Theta}|\mathbf{p}|\,,\quad\breve{\Theta}=\theta\sqrt{1+\beta^2[\beta^2-2\sin(2\phi)]\sin^2\vartheta}\,,
\end{equation}%
so the corresponding mode propagates as a Podolsky-type mode with a modified mass depending on the
momentum direction. Note the similarities to Eq.~\eqref{eq:group-velocity-parity-even-small-momentum}.
The limit $\beta\mapsto 0$ for vanishing Lorentz violation reproduces the behavior of the conventional Podolsky mode.

For this spacelike configuration, we now investigate the dispersion relation derived from
Eq.~(\ref{PDR6B}) by initially studying the case of orthogonal vectors
$\mathbf{B}$, $\mathbf{C}$. This configuration yields two different additional dispersion relations, written as
\begin{subequations}
\begin{equation}
(p_0^{(\pm)})^2=\frac{\mathbf{p}^2}{z^4-x^4}\left\{\mathcal{V}/2
\pm\sqrt{(z^4-x^4){\left[x^4\cos^2\vartheta-\mathcal{W}\right]}
+(\mathcal{V}/2)^2}\right\}\,,
\label{DR44}%
\end{equation}
where
\begin{align}
\mathcal{V} &  =2z^2(1+z^2)-2x^4+\sin^2\vartheta\left[x^2+(x^2-2y^2)z^2\sin(2\phi)\right]\,, \\[2ex]
\mathcal{W} &  =\left[1+z^2-y^2\sin^{2}\vartheta\sin(2\phi)\right]\left[1+z^2+(x^2-y^2)\sin^2\vartheta\sin(2\phi)\right]\,,
\end{align}
and%
\begin{equation}
x=\sqrt{|\mathbf{B}||\mathbf{C}|}\eta_{1}|\mathbf{p}|\,,\quad
y=\sqrt{|\mathbf{B}||\mathbf{C}|}\eta_{2}|\mathbf{p}|\,,\quad z=\theta|\mathbf{p}|\,.
\label{DEFIN}%
\end{equation}
The modulus squared of the group velocity is given by a highly involved expression
\end{subequations}
\begin{subequations}
\label{eq:group-velocity-full-theory-complicated-dispersion-equation-spacelike-orthogonal}%
\begin{equation}
(u_{\mathrm{gr}}^{(\pm)})^2=\frac{|\mathbf{p}|^2}{(p_0^{(\pm)})^2}\left(
\cos^2\vartheta+\frac{{f_{\pm}^2}(\vartheta,\phi)+{f_{\pm}^2}%
(\vartheta,\pi/2-\phi)}{{g_{\pm}^2}(\vartheta,\phi)}\right)\,,
\end{equation}
with
\begin{align}
f_{\pm}(\vartheta,\phi) &  =2y^2(y^2-x^2)\sin^3\vartheta\sin(2\phi)\sin\phi\nonumber\\
&\phantom{{}={}}+(x^2-2y^2)\left\{1+z^2\left[2\sin^2\vartheta\cos^2\phi+1-(e^{(\pm)})^{2}\right]\right\}\sin\vartheta\sin\phi\nonumber\\
&\phantom{{}={}}+\left(x^4\left[(e^{(\pm)})^2-\cos^2\vartheta\right]
+2z^2\left\{1+\left[1-(e^{(\pm)})^{2}\right]z^2\right\}\right)\sin\vartheta\cos\phi\,,\displaybreak[0]\\[2ex]
g_{\pm}(\vartheta,\phi) &  =\left[2(e^{(\pm)})^2-1-\cos^2\vartheta\right]x^4+\left[2+(x^2-2y^2)\sin^2\vartheta\sin(2\phi)\right]z^2\nonumber\\
&\phantom{{}={}}+2\left[1-(e^{(\pm)})^{2}\right]z^4\,,
\end{align}
\begin{figure}[ptb]
\centering
\subfloat[]{\label{fig:full-spacelike-orthogonal-case-1}\includegraphics[scale=0.35]{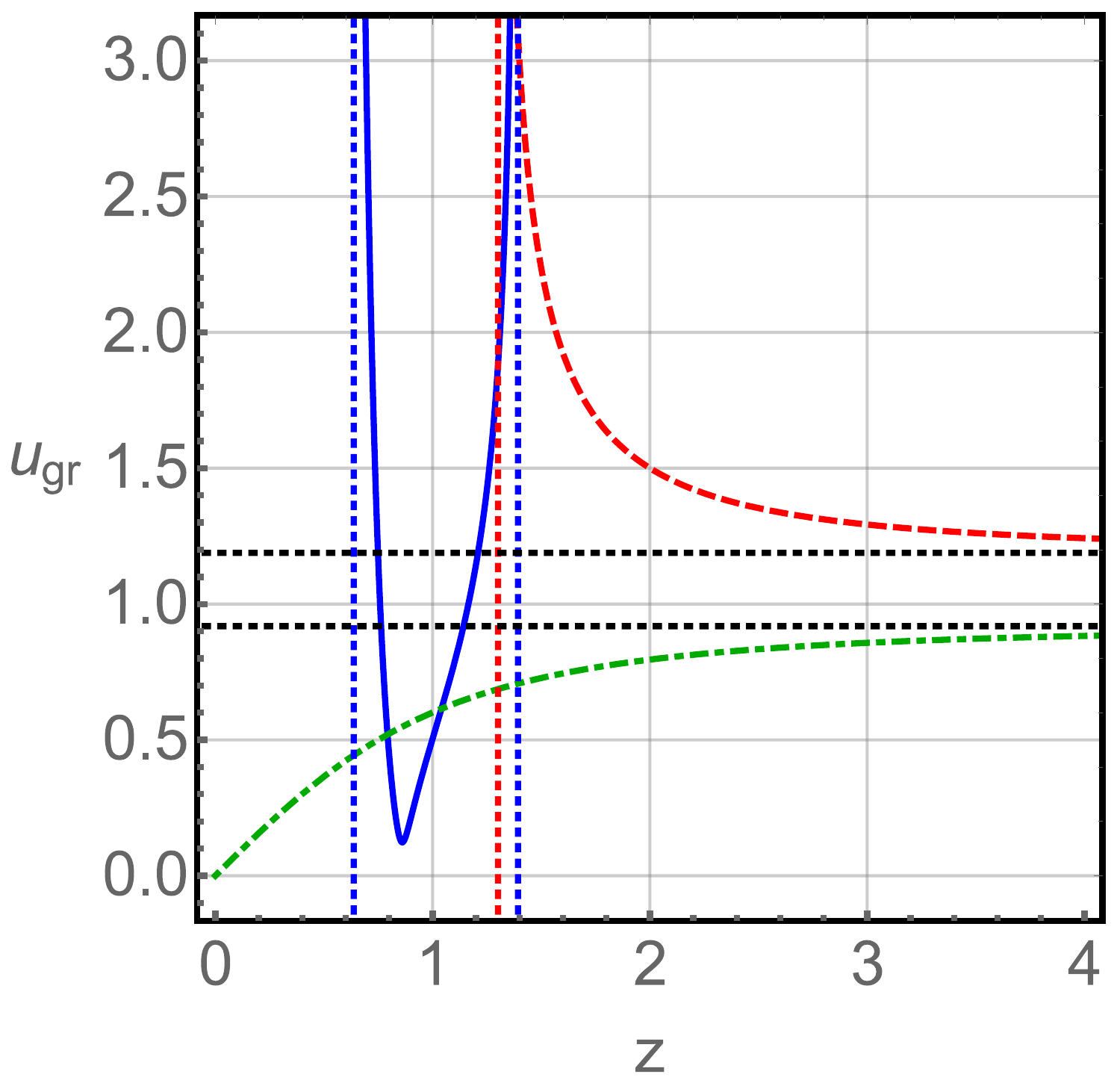}}\hspace
{2cm}
\subfloat[]{\label{fig:full-spacelike-orthogonal-case-2}\includegraphics[scale=0.35]{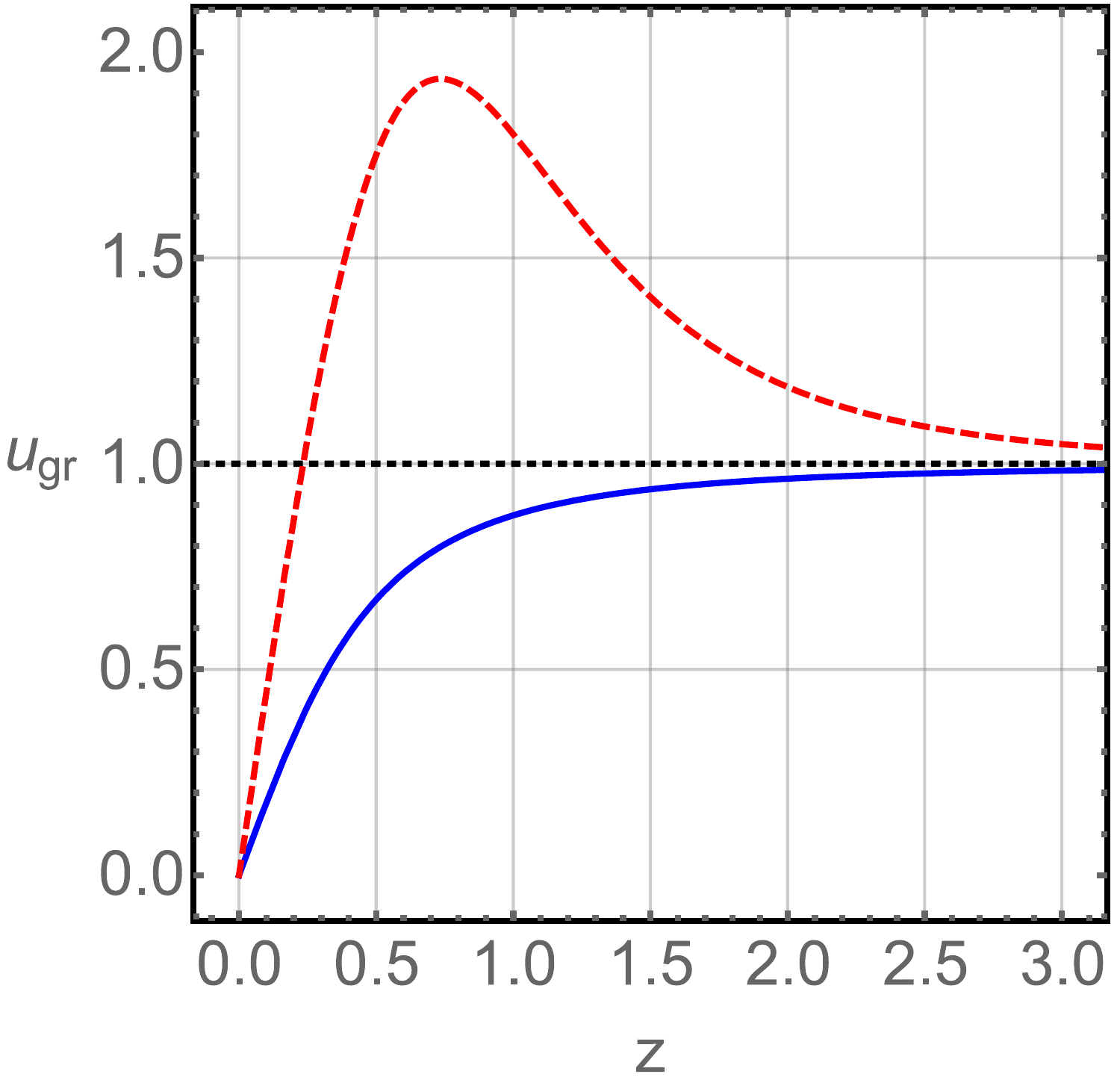}}
\caption{Group velocity of
Eq.~\eqref{eq:group-velocity-full-theory-complicated-dispersion-equation-spacelike-orthogonal}
of the mode $\oplus$ for the parameters $\alpha=3/2$, $\beta=2$,
$\vartheta=\pi/2$, $\phi=\pi/3$ (blue, plain), for $\alpha=3/2$, $\beta=1$,
$\vartheta=\pi/4$, $\phi=0$ (red, dotted), and for $\alpha=1/3$, $\beta=1/2$,
$\vartheta=\pi/2$, $\phi=\pi/4$ (green, dash-dotted)
\protect\subref{fig:full-spacelike-orthogonal-case-1}. The group velocity of the mode
$\ominus$ for $\alpha=3/2$, $\beta=1$, $\vartheta=0$, $\phi=0$ (blue, plain)
and for $\alpha=3/2$, $\beta=3$, $\vartheta=\pi/2$, $\phi=0$ (red, dotted) is
shown in \protect\subref{fig:full-spacelike-orthogonal-case-2}. Vertical asymptotes at
singularities are illustrated via color-coded, dotted lines, whereas horizontal asymptotes
are shown as black, dotted lines.}%
\label{caseOrtog2}%
\end{figure}
where $e^{(\pm)}\equiv p_0^{(\pm)}/|\mathbf{p}|$.
Fig.~\ref{caseOrtog2} shows the group velocity for the modes $\oplus$
and $\ominus$ for different angles and parameters $\sqrt{|\mathbf{B}%
||\mathbf{C}|}\eta_1/\theta\equiv\alpha$ and $\sqrt{|\mathbf{B}%
||\mathbf{C}|}\eta_2/\theta\equiv\beta$. The group velocity of the mode
$\oplus$ is badly behaved for a broad range of parameters and angles, as it
exhibits one or even two singularities that lie at the following values of
$z$:
\end{subequations}
\begin{equation}
z_{\mathrm{sing}}^{(\pm)}=\frac{\sqrt{2}}{\sqrt{(2\beta^2-\alpha^2)
\sin^2\vartheta\sin(2\phi)-2\pm\alpha^2\sqrt{4\cos^2\vartheta+\sin^4
\vartheta\sin^2(2\phi)}}}\,.
\end{equation}
Besides, when there are singularities, the group velocity
even becomes complex for large domains of the momentum. For small momenta, the
group velocity for both modes exhibits the following asymptotic behavior:
\begin{subequations}
\begin{align}
u_{\mathrm{gr}}^{(\pm)}&=\bar{\Theta}^{(\pm)}|\mathbf{p}|\,, \\[2ex]
\bar{\Theta}^{(\pm)}&=\theta\sqrt{1\mp\alpha^2+\frac{\sin^2\vartheta\left[\alpha^2(\pm\alpha^2-2)\pm 2\beta^2(\alpha^2-\beta^2)+(\alpha^2\mp 2)(\alpha^2-2\beta^2)\sin(2\phi)\right]}{2(\alpha^2\mp1)}}\,.
\end{align}
\end{subequations}
Hence, these modes are modifications of the Podolsky mode with anisotropic
Podolsky parameters that can become complex when the parameters are chosen suitably.
The Podolsky mode is reproduced in the combined limit $\alpha\mapsto 0$ and $\beta\mapsto 0$.
For certain choices, the group velocity is well-behaved in the sense that it increases
monotonically from $0$ until it approaches a value smaller than 1 from below, cf. the green
curve in Fig.~\ref{fig:full-spacelike-orthogonal-case-1}. It then behaves like an exotic mode.
The other curves in this figure correspond to parameter choices that provide spurious behaviors.
In contrast, the mode $\ominus$ in Fig.~\ref{fig:full-spacelike-orthogonal-case-2} does not
exhibit any singularities. It rises from 0 and converges to a value smaller then 1 from below
or larger than 1 from above. The latter behavior, though, is interpreted as a violation of
causality, which occurs at least for some parameter choices. Therefore, the mode represented by
the blue curve is exotic, whereas the mode illustrated by the red curve is spurious.

We can also present expansions for the front velocities
following from DR~(\ref{DR44}) in the limit of a large Podolsky
parameter $\theta$ in comparison to the LV coefficients
$\eta_{1}$, $\eta_{2}$:
\begin{subequations}
\begin{align}
u_{\mathrm{fr}}^{(+)} &  \approx1+\frac{1}{2}\left[\left(\frac{x}{z}\right)^2
-\left(\frac{y}{z}\right)^2\right]\sin^2\vartheta\sin(2\phi)\,,\\[2ex]
u_{\mathrm{fr}}^{(-)} &  \approx1-\frac{1}{2}\left(\frac{y}{z}\right)^2\sin^2\vartheta\sin(2\phi)\,.
\end{align}
Both can be larger than 1, depending on the relative sizes of $x$,
$y$ and on the angle $\phi,$ which spoils the premises of causality.
Thus, while the mode $\oplus$ is  unphysical in several
senses, the mode $\ominus$ breaks causality for
certain choices of parameters.

The other configuration to be examined is the case of parallel
vectors $\mathbf{B}$, $\mathbf{C}$ where the dispersion equation~(\ref{PDR6B})
yields%
\end{subequations}
\begin{subequations}
\begin{align}
E^{(\pm)} &  =\zeta^{(\pm)}|\mathbf{p}|\,,\quad{\zeta^{(\pm)}=\zeta^{(\pm)}(|\mathbf{p}|)}\,,\displaybreak[0]\\[2ex]
\zeta^{(+)} &  =\frac{\sqrt{1+z^2-2y^2\cos^2\vartheta}}{z}\,,\displaybreak[0]\\[2ex]
\zeta^{(-)} &  =\sqrt{\frac{1+z^2-2(x^2\sin^2\vartheta+y^2\cos^2\vartheta)}{z^2-2x^2}}\,.
\end{align}
The front velocities of both modes are given by
\end{subequations}
\begin{subequations}
\begin{align}
u_{\mathrm{fr}}^{(+)} &  =\sqrt{1-2\beta^2\cos^2\vartheta}%
\,,\displaybreak[0]\\[2ex]
u_{\mathrm{fr}}^{(-)} &  =\sqrt{\frac{1-2(\alpha^2\sin^2\vartheta
+\beta^2\cos^2\vartheta)}{1-2\alpha^2}}\,,
\end{align}
\begin{figure}[ptb]
\centering
\subfloat[]{\label{fig:full-theory-complicated-dispersion-equation-spacelike-parallel-case-1}
\includegraphics[scale=0.35]{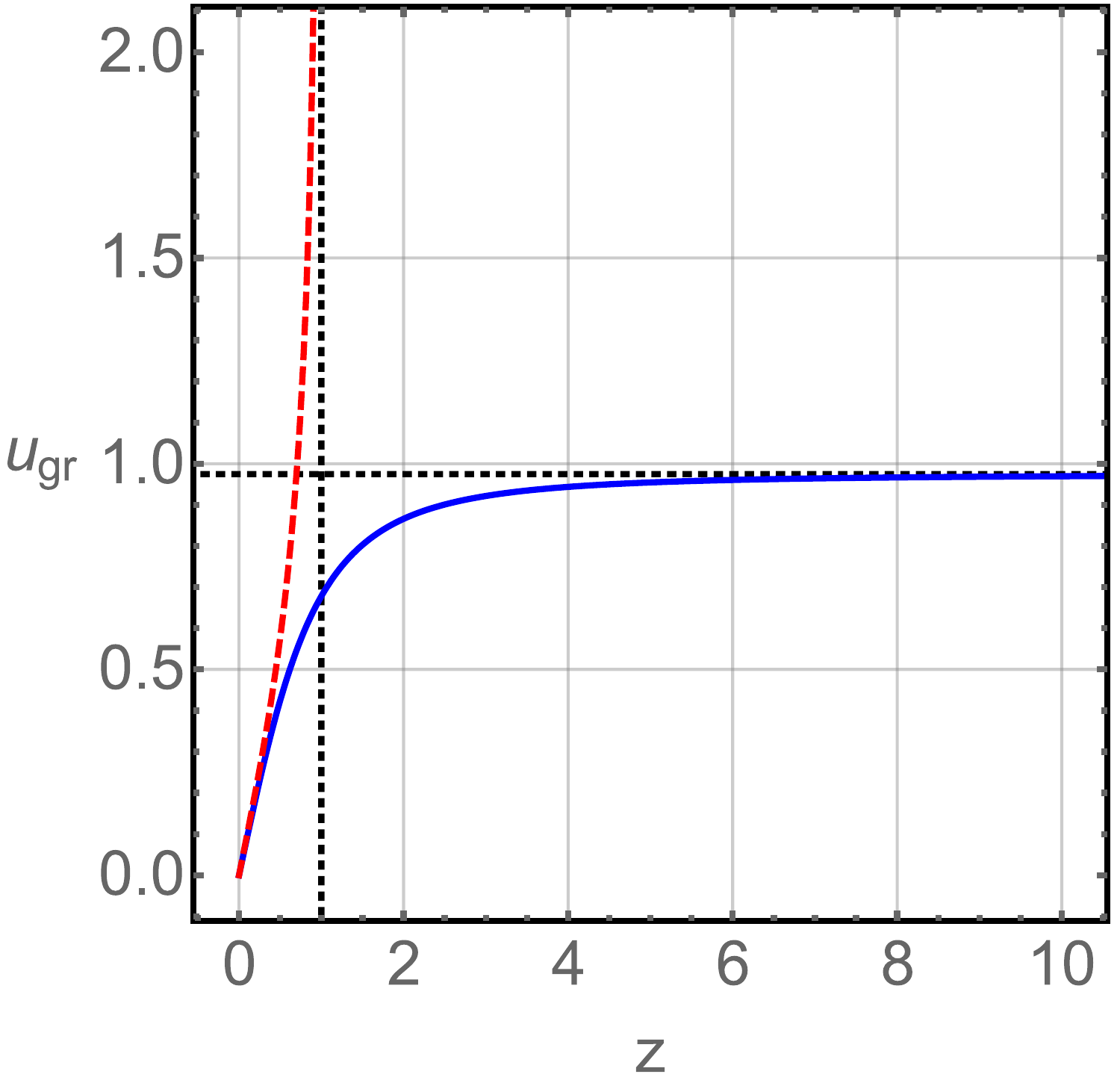}}\hspace{2cm}
\subfloat[]{\label{fig:full-theory-complicated-dispersion-equation-spacelike-parallel-case-2}
\includegraphics[scale=0.35]{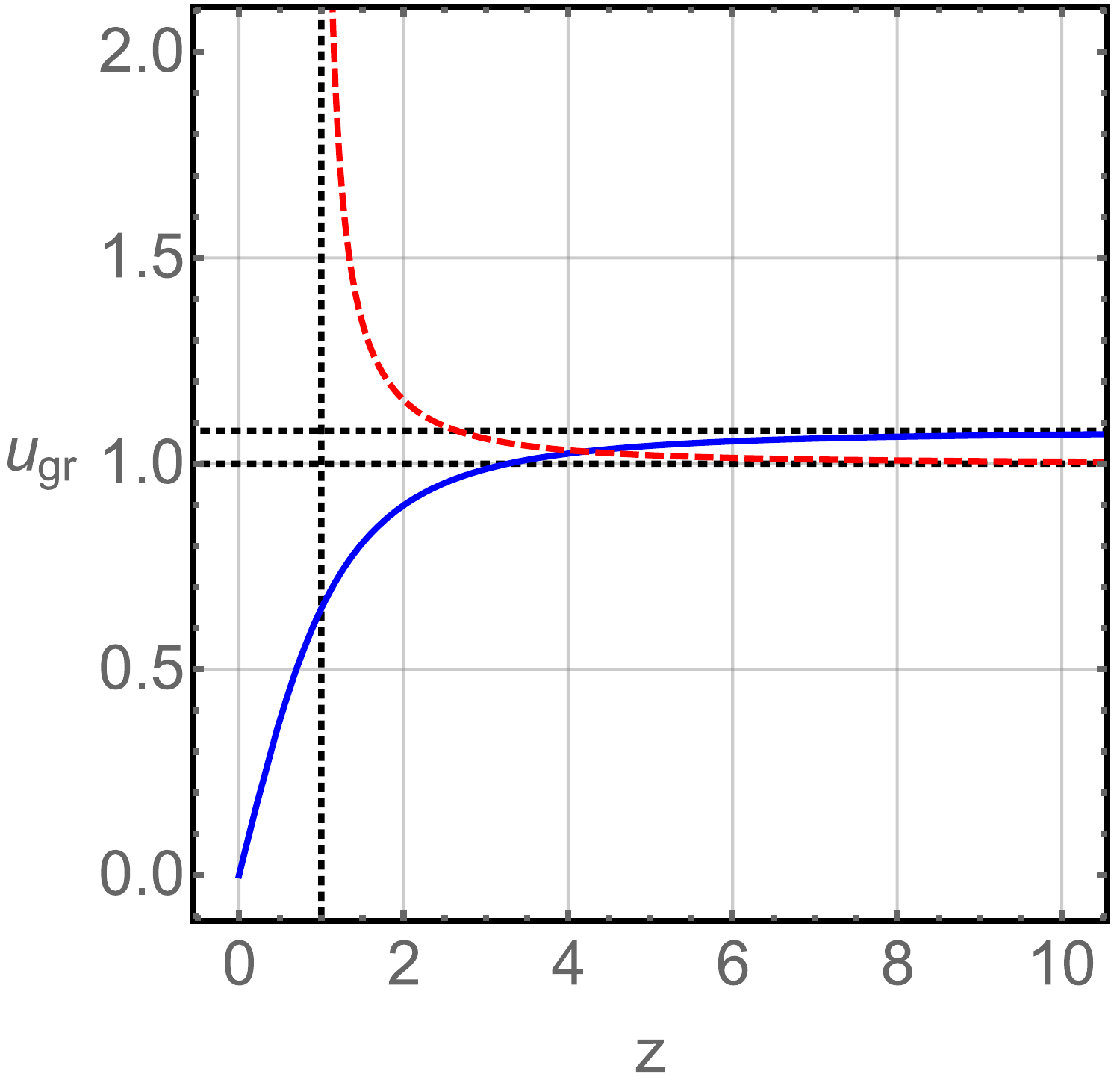}}
\caption{Group velocity of Eq.~\eqref{eq:group-velocity-spacelike-case-parallel-vectors}
of the mode $\oplus$ for the parameters $\beta=1/(2\sqrt{2})$, $\vartheta=\pi/3$ (blue, plain) and for $\beta=1$, $\vartheta=0$ (red, dotted) \protect\subref{fig:full-theory-complicated-dispersion-equation-spacelike-parallel-case-1}. The group velocity of the mode $\ominus$ for
the parameter choices $\alpha=1/2$, $\beta=1/(2\sqrt{2})$, $\vartheta=\pi/3$ (plain, blue) and $\alpha=\beta=1$, $\vartheta=0$ (red, dotted)
can be found in \protect\subref{fig:full-theory-complicated-dispersion-equation-spacelike-parallel-case-2}. Asymptotes are shown as black,
dotted lines.}
\label{caseN6}%
\end{figure}
where these results are expressed in terms of the constant ratios
$x/z\equiv\alpha$ and $y/z\equiv\beta$. The front velocity of the mode
$\oplus$ cannot exceed 1, which does not hold for the mode $\ominus$, though,
as $u_{\mathrm{fr}}^{(-)}$ can be larger than 1 for $\beta<\alpha$.
The next step is to obtain the magnitudes of the group velocities:
\end{subequations}
\begin{subequations}
\label{eq:group-velocity-spacelike-case-parallel-vectors}%
\begin{align}
u_{\mathrm{gr}}^{(+)} &  =\sqrt{\frac{4y^2(y^2-z^2)\cos^2%
\vartheta+z^4}{z^2(1+z^2-2y^2\cos^2\vartheta)}}\,,\displaybreak[0]\\[2ex]
u_{\mathrm{gr}}^{(-)} &  =\sqrt{\frac{4(x^2-y^2)(x^2+y^2-z^2)\cos^2\vartheta
-(2x^2-z^2)^{2}}{(2x^2-z^2)\left[1+z^2-2(x^2\sin^2\vartheta+y^2\cos^2\vartheta)\right]}}\,,
\end{align}
\end{subequations}
with the definitions (\ref{DEFIN}). By introducing the parameters
$\alpha$, $\beta$ into the previous results, the asymptotic behavior of the group
velocity for small momenta is
\begin{subequations}
\begin{align}
u_{\mathrm{gr}}^{(\pm)}&=\bar{\bar{\Theta}}^{(\pm)}|\mathbf{p}|\,, \\[2ex]
\bar{\bar{\Theta}}^{(+)}&=\theta\sqrt{1+4\beta^2(\beta^2-1)\cos^2\vartheta}\,,\quad\bar{\bar{\Theta}}^{(-)}=\theta\sqrt{1-2\alpha^2+\frac{4(\alpha^2-\beta^2)[1-(\alpha^2+\beta^2)]\cos^2\vartheta}{1-2\alpha^2}}\,.
\end{align}
\end{subequations}
Each mode again resembles an anisotropic Podolsky mode. The limit of vanishing
Lorentz violation leads to the standard Podolsky dispersion relation.
Also, the modes can exhibit spurious behavior, since singularities are present for
\begin{equation}
z_{\mathrm{sing}}^{(+)}=\frac{1}{\sqrt{2\beta^2\cos^2\vartheta-1}}\,,\quad
z_{\mathrm{sing}}^{(-)}=\frac{1}{\sqrt{2(\alpha^2\sin^2\vartheta+\beta^2\cos^2\vartheta)-1}}\,,
\end{equation}
again pointing to the existence of unphysical regimes (see the red curves of Fig.~\ref{caseN6}).
The parameters $\alpha$ and $\beta$ can be chosen such that these values are
complex, whereupon there are no singularities. In this case, the group
velocity rises from 0 to a finite value. This behavior is
characteristic of an exotic mode if this finite value is smaller than 1. The blue curve in Fig.~\ref{fig:full-theory-complicated-dispersion-equation-spacelike-parallel-case-1} represents an example for this case.
If the finite value is larger than 1 the mode is spurious whereby an example is given by the blue curve in Fig.~\ref{fig:full-theory-complicated-dispersion-equation-spacelike-parallel-case-2}.
Each expression for the group velocity exceeds 1 for
\begin{equation}
z|_{u_{\mathrm{gr}}^{(+)}=1}=\frac{1}{\sqrt{4\beta^2-2}|\beta\cos\vartheta|}\,,\quad
z|_{u_{\mathrm{gr}}^{(-)}=1}=\sqrt{\frac{1-2\alpha^2}{(1-2\beta^2)(\alpha^2-\beta^2)[1+\cos(2\vartheta)]}}\,.
\end{equation}
The first value becomes complex for $\beta<1/\sqrt{2}$ and the second for
$\alpha>1/\sqrt{2}$, $\beta>1/\sqrt{2}$, and $\beta>\alpha$ at the same time
or for $\alpha>1/\sqrt{2}>\beta$ or when $1/\sqrt{2}>\beta>\alpha$. Under these
conditions, the group velocity stays below 1.
The profiles of $u_{\mathrm{gr}}$ for these two modes are depicted in
Fig.~\ref{caseN6}.

\section{Connection to nonminimal SME}

An interesting question to ask is whether the LV contribution of
Lagrangian~(\ref{LVSLAD1}) is contained in the nonminimal SME by
Kosteleck\'{y} and Mewes \cite{Kostelec1}. As we are addressing LV and
\textit{CPT}-even dimension-6 terms, the first higher-order LV coefficient
to be considered is that of dimension 6, namely%
\begin{equation}
(\hat{k}_{F})^{\kappa\lambda\mu\nu}=
(k_{F}^{(6)})^{\kappa\lambda\mu\nu\alpha_1\alpha_2}\partial_{\alpha_1}\partial_{\alpha_2}\,,
\label{k6AM}%
\end{equation}
where we have omitted the usual dimension-4 \textit{CPT}-even coefficients of the
photon sector of the SME, $(k_F^{(4)})^{\kappa\lambda\mu\nu}$.
It is possible to show that the parameterization%
\begin{subequations}
\label{eq:preliminary-mappings}
\begin{equation}
(k_{F}^{(6)})^{\mu\nu\varrho\sigma\alpha_1\alpha_2}=\eta^{2}D^{\nu\sigma}
g^{\mu\alpha_1}g^{\varrho\alpha_2}\,,%
\end{equation}
successfully reproduces the LV term present in Lagrangian~(\ref{LVSLAD1}).
Indeed,%
\begin{equation}
(k_{F}^{(6)})^{\mu\nu\varrho\sigma\alpha_1\alpha_2}\partial_{\alpha_1}
\partial_{\alpha_2}F_{\mu\nu}F_{\rho\sigma}=\eta^2D^{\nu\sigma}%
\partial^{\mu}\,F_{\mu\nu}\partial^{\varrho}F_{\rho\sigma}\,,
\end{equation}
\end{subequations}
which demonstrates that this LV piece is contained in the nonminimal SME~\cite{Kostelec1}. In an analogue way,
the correspondence
\begin{equation}
(k_F^{(6)})^{\mu\nu\varrho\sigma\alpha_1\alpha_2}=
\eta_1^2D^{\nu\sigma}g^{\mu\alpha_1}g^{\varrho\alpha_2}
-\eta_2^2D^{\alpha_2\sigma}g^{\mu\alpha_1}g^{\nu\varrho}\,,
\label{Eq-88}
\end{equation}
yields the LV terms of the generalized model of Lagrangian (\ref{L3}). Finally, the symmetries of $(\hat{k}_F)^{\kappa\lambda\mu\nu}$ have to be taken into account. The latter fourth-rank observer tensor is antisymmetric
in the first and second pair of indices and symmetric under an exchange of both index pairs, which is evidently not the case
for the choices in Eqs. (\ref{eq:preliminary-mappings}), (\ref{Eq-88}). Performing the (anti)symmetrization explicitly
and considering the factor 1/4 before $(\hat{k}_F)^{\kappa\lambda\mu\nu}$ in the definition of the $CPT$-even term of the SME
photon sector, the final mapping of our terms to the SME is given by
\begin{eqnarray}
(\hat{k}_F)^{\mu\nu\varrho\sigma}&=&\frac{1}{2}\left[(k_F^{(6)})^{\mu\nu\varrho\sigma\alpha_1 \alpha_2}-(k_F^{(6)})^{\nu\mu\varrho\sigma\alpha_1\alpha_2}-(k_F^{(6)})^{\mu\nu\sigma\varrho \alpha_1\alpha_2}+(k_F^{(6)})^{\nu\mu\sigma\varrho\alpha_1\alpha_2}\right. \notag \\
&&\left.\phantom{\frac{1}{2}\Big[}+\,(\mu\leftrightarrow\varrho,\nu\leftrightarrow \sigma)\right]\partial_{\alpha_1}\partial_{\alpha_2}\,.
\end{eqnarray}
It can be checked that the tensor constructed in this way automatically satisfies the Bianchi identity for
an arbitrary triple of indices chosen from the first four indices. In addition, antisymmetrization on any three indices must be
imposed via Eq.~(55) of the first paper in \cite{Kostelec1}.
The Podolsky term is Lorentz-invariant and cannot be mapped to SME coefficients. However, it is introduced into the field equations
of the nonminimal $CPT$-even photon sector as follows:
\begin{subequations}
\begin{align}
M^{\mu\nu}A_{\nu}&=0\,, \\[2ex]
M^{\mu\nu}&=\left\{(1-\theta^2p^2)\left[g^{\mu\nu}g^{\alpha\beta}-g^{\mu\alpha} g^{\nu\beta}\right]+2(\hat{k}_F)^{\mu\alpha\nu\beta}\right\}p_{\alpha}p_{\beta}\,.
\end{align}
\end{subequations}

\section{Conclusion and remarks}

In this work we have addressed an electrodynamics endowed with LV terms of mass dimension 6.
We considered aspects not yet analysed in the literature, such as an evaluation of the
propagator, the dispersion relations, and the propagation modes. First, a Maxwell electrodynamics
modified by an anisotropic Podolsky-type term of mass dimension 6 was examined. For certain parameter
choices, all the modes, which are affected by the higher-dimensional term, were found to exhibit a
nonphysical behavior. The magnitude of the group velocity was demonstrated to become larger than 1
(breaking of causality), to have singularities or to vanish for small momenta (absence of
propagation). The first two behaviors described are interpreted as unphysical and
we referred to dispersion relations having characteristics of this kind as spurious.
Features such as decreasing group or front velocities for small momenta only indicate
that these modes decouple from the low-energy regime. According to the criteria used in
the paper, a mode endowed with such properties was called exotic as long as it neither
involved singularities nor superluminal group or front velocities.

The results obtained indicate that an electrodynamics modified by this kind of higher-derivative
LV term can only provide a physical propagation of signals in the limit of large momenta. Such
an interpretation could recover a meaningful behavior of the theory in the case that it is
possible to state a suitable cutoff without generating bad side effects. For small momenta, some
of the modes correspond to Lorentz-violating modifications of the Podolsky-type dispersion
relation, which are not necessarily unphysical. We also found dispersion relations
providing group velocities that are larger than 1 or divergent at some points, which are
behaviors characterizing spurious propagation modes. A brief discussion of unitarity for this
modification was delivered as well. The result of our analysis was that additional time
derivatives in the Lagrange density are likely to spoil unitarity. This behavior is
expected and was already observed in Lorentz-invariant higher-derivative theories as well as
alternative LV higher-derivative modifications of the photon and fermion sector.

At a second stage, we analyzed a more general extension of the modified electrodynamics in the
presence of an additional anisotropic Lee-Wick term of mass dimension 6. Again, the propagator was
derived and the corresponding dispersion relations were obtained from it. Subsequently, these
dispersion relations were examined with respect to signal propagation. The structure of this second
framework is more involved in comparison to the first theory under consideration. The modified
dispersion relations are associated with exotic and spurious modes, as already observed in the first
model. There exist parameter choices that are connected to issues with causality. Another observation
was that some of the dispersion relations can exhibit singularities, i.e., they become unphysical for a
certain range of momenta, which is enough to classify these modes as spurious. However, we also
found sectors for which the dispersion laws do not suffer from any of the problems just mentioned.
Therefore, those can be considered as well-behaved according to the criteria that the current
paper is based on.

The results of the paper suggest which sectors of the framework should be subject
to quantization and which ones should be discarded. These sectors are classified according
to the geometrical properties of the two four-vectors that the background tensor is composed of.
Both spurious and exotic modes were found dependent on choices of the parameters.
There is no need of introducing lower cutoffs of the four-momentum, which would be difficult to
realize from a physical point of view, anyhow. Instead, the regions of parameter space associated
with a vicious classical behavior should be removed from the theory. If the theories proposed are subject
to studies at the quantum level in the future,
the results of the paper demonstrate, which sectors could be quantized properly. Physical dispersion
relations must be clearly free of singularities. Modes without violations of classical causality may
be the preferred ones to be studied. However, it ought to be taken into account, as well, that
superluminal velocities do not necessarily cause problems at the quantum level, e.g., for microcausality
\cite{Klinkhamer:2010zs}.

One of the most significant questions to be answered in future works would be related to gauge fixing.
It was mentioned in the paper that gauge fixing in vector field theories including higher derivatives
is even more subtle than in the standard case. Whether or not the gauge fixing conditions proposed
in \cite{GalvaoP} for Podolsky's theory are still applicable in the Lorentz-violating case, would be an
interesting aspect to investigate. Furthermore, to quantize the theory, proper field operators will have to be
constructed and a treatment of the spurious dispersion laws at the quantum level would be mandatory.
A related question is whether the prescription of how to preserve (perturbative) unitarity,
which was introduced by Lee and Wick \cite{LeeWick}, still works. Finally, the free theory should be
coupled to fermions to be able to study physical processes. Such questions have been considered recently for
a particular fermionic framework including higher derivatives \cite{Reyes:2016pus}, but how to proceed in the photon sector
is still an open problem.

\begin{acknowledgments}
The authors are grateful to FAPEMA, CAPES, and CNPq (Brazilian research
agencies) for invaluable financial support.
\end{acknowledgments}

\end{document}